\documentclass[journal]{IEEEtran}

\makeatletter
\def\ps@headings{%
\def\@oddhead{\mbox{}\scriptsize\rightmark \hfil \thepage}%
\def\@evenhead{\scriptsize\thepage \hfil \leftmark\mbox{}}%
\def\@oddfoot{}%
\def\@evenfoot{}}

\makeatother
\usepackage[english]{babel}
\usepackage[utf8]{inputenc}
\usepackage{amsthm}
\usepackage{amsmath,amssymb}
\usepackage{pifont}

\usepackage{makecell}
\usepackage{algorithm, algcompatible}

\usepackage{bbold}
\usepackage[usenames,dvipsnames]{color}
\usepackage{float}
\usepackage{cite}

\usepackage{tikz}
\usetikzlibrary{shapes.geometric, arrows, calc}

\usepackage{comment}

\newtheorem{definition}{\hskip 0pt Definition}
\newtheorem{theorem}{\hskip 0pt Theorem}
\newtheorem{lemma}{\hskip 0pt Lemma}

\usepackage{graphicx}

\newcommand{\dm}[0]{\Delta_{\min}}

\floatstyle{ruled}
\newfloat{algorithm}{tbp}{loa}
\providecommand{\algorithmname}{Algorithm}
\floatname{algorithm}{\protect\algorithmname}

\DeclareMathOperator{\Ep}{\mathop{\mathbb{E}}}

\newcommand{\va}{\mbox{${\bf a}$}}

\newcommand{\vr}{\mbox{${\bf r}$}}
\newcommand{\vv}{\mbox{${\bf v}$}}

\newcommand{\vzero}{\mbox{${\bf 0}$}}

\newcommand{\ga}{\alpha}
\newcommand{\gb}{\beta}

\newcommand{\gd}{\delta}

\newcommand{\gl}{\lambda}
\newcommand{\gm}{\mu}

\newcommand{\gr}{\rho}

\newcommand{\gt}{\tau}

\newcommand{\gD}{\Delta}

\newcommand{\cA}{{\cal A}}

\newcommand{\cK}{{\cal K}}

\newcommand{\cM}{{\cal M}}
\newcommand{\cN}{{\cal N}}

\newtheorem{question}{Question}[section]
\newtheorem{coro}{Corollary}[section]

\newcommand{\beq}{\begin{equation}}
\newcommand{\eeq}{\end{equation}}
\newcommand{\bea}{\begin{array}}
\newcommand{\ena}{\end{array}}
\newcommand{\bds}{\begin {itemize}}
\newcommand{\eds}{\end {itemize}}
\newcommand{\bdf}{\begin{definition}}
\newcommand{\blm}{\begin{lemma}}
\newcommand{\edf}{\end{definition}}
\newcommand{\elm}{\end{lemma}}
\newcommand{\bthm}{\begin{theorem}}
\newcommand{\ethm}{\end{theorem}}
\newcommand{\bprp}{\begin{prop}}
\newcommand{\eprp}{\end{prop}}
\newcommand{\bcl}{\begin{claim}}
\newcommand{\ecl}{\end{claim}}
\newcommand{\bcr}{\begin{coro}}
\newcommand{\ecr}{\end{coro}}
\newcommand{\bquest}{\begin{question}}
\newcommand{\equest}{\end{question}}

\makeatletter
\newcommand{\pushright}[1]{\ifmeasuring@#1\else\omit\hfill$\displaystyle#1$\fi\ignorespaces}
\newcommand{\pushleft}[1]{\ifmeasuring@#1\else\omit$\displaystyle#1$\hfill\fi\ignorespaces}
\makeatother
\begin{document}

\title{Distributed Learning for Optimal Spectrum Access in Dense Device-to-Device Ad-Hoc Networks}

\author{Tomer~Boyarski, Wenbo~Wang,~\IEEEmembership{Senior Member,~IEEE} and
Amir~Leshem,~\IEEEmembership{Fellow,~IEEE}
\thanks{
Tomer~Boyarski and Amir~Leshem are with the Faculty of Engineering, Bar-Ilan University, Ramat Gan 5290002, Israel. (e-mails: tomer.boyarski@gmail.com; amir.leshem@biu.ac.il)
Wenbo Wang is with the Faculty of Mechanical and Electrical Engineering, Kunming University of Science and Technology (KUST), Kunming 650500, China (e-mail: wenbo\_wang@kust.edu.cn). The research was partially funded by ISF grant 1644/18.
 }\vspace{-5mm}}
\maketitle
\begin{abstract}
  In 5G networks, Device-to-Device (D2D) communications aim to provide dense coverage without relying on the cellular network infrastructure. To achieve this goal, the D2D links are expected to be capable of self-organizing and allocating finite, interfering resources with limited inter-link coordination. We consider a dense ad-hoc D2D network and propose a decentralized time-frequency allocation mechanism that achieves sub-linear social regret toward optimal spectrum efficiency. The proposed mechanism is constructed in the framework of multi-agent multi-armed bandits, which employs the carrier-sensing-based distributed auction to learn the optimal allocation of time-frequency blocks with different channel state dynamics from scratch. Our theoretical analysis shows that the proposed fully distributed mechanism achieves a logarithmic regret bound by adopting an epoch-based strategy-learning scheme when the length of the strategy-exploitation window is exponentially growing. We further propose an implementation-friendly protocol featuring a fixed exploitation window, which guarantees a good tradeoff between performance optimality and protocol efficiency. Numerical simulations demonstrate that the proposed protocol achieves higher efficiency than the prevalent reference algorithms in both static and dynamic wireless environments.
\end{abstract}

\section{Introduction}
The rapid development of 5G technologies has triggered an unprecedented demand for high-throughput and ubiquitous connectivity, especially among various machine-type terminals. However, as more wireless devices are densely deployed, and the connections become dominated by machine terminals in proximity, infrastructure-based communications face the problems of heavy loads on the base station side and weak scalability. As a result, Device-to-Device (D2D) networks, with their intrinsic characteristics of self-organization through direct connections and ability to heterogeneously operate over the existing network infrastructure (e.g., over unlicensed bands) are considered a promising and cost-efficient solution for improving spectrum utilization as a complement to next-generation cellular networks~\cite{8340813,6231164}.

Self-organization can significantly reduce the signaling overheads and allow better scalability in a D2D network. However, unlike the cellular network, there is no central controller in a D2D network to collect global Channel State Information (CSI). Thus, no centralized spectrum resource optimization is directly applicable. In addition, the uncontrolled nature of the unlicensed spectrum presents a significant challenge to the ad-hoc devices such that local cognition of the dynamic channel states becomes necessary to guarantee QoS provision. As a result, it is of utmost importance to design a suitable distributed mechanism for resource allocation that achieves high spectrum efficiency over channels of heterogeneous quality while not overwhelming the signaling bottleneck of the D2D links due to the nature of the ad-hoc connections.

Learning optimal allocations distributedly needs to exploit the properties of the wireless medium to reduce signaling overhead, which becomes prohibitively complex in large-scale networks. In this paper, we consider a dense D2D ad-hoc wireless network where the number of D2D links can be much larger than that of available channels. For such a scenario, we propose a fully distributed time-frequency resource allocation protocol for medium access, which exploits standard random access protocols (e.g., CSMA) to distributedly achieve an optimal orthogonal allocation with minimal overhead.
To theoretically guarantee the optimality of the proposed scheme, we present the distributed protocol as a solution to a multi-agent Multi-Armed Bandits (MAB) problem. Our proposed mechanism employs the distributed auction technique as in~\cite{zafaruddin2019distributed} for socially optimal resource allocation. Compared to~\cite{zafaruddin2019distributed}, the proposed technique contains several fundamental improvements. First, we demonstrate that~\cite{zafaruddin2019distributed} can be generalized into an OFDMA-like allocation. To accommodate the time-frequency allocation, a more complex framing structure is proposed. Second, for the practical protocol implementation, a carrier-sensing-based distributed auction mechanism with fixed quantization accuracy is designed to ensure the completion of coordination with a much shorter contention window. Furthermore, our scheme requires a fixed-size frame for strategy exploitation, and thus is more convenient for implementation.

The rest of this paper is organized as follows. Section~\ref{sec:survey} provides a brief review of related works. Section~\ref{sec:system} presents the problem formulation. Section~\ref{sec:protocol} proposes a novel distributed learning-based scheme for channel allocation, and Section \ref{sec:regret} presents a theoretical analysis of the regret of the proposed learning algorithm. Section~\ref{sec:practical} addresses the implementation issues related to the MAC protocol design based on our proposed learning algorithm. Section~\ref{sec:simulations} demonstrates the efficiency of the proposed scheme with numerical simulations. Finally, Section~\ref{sec:conclude} concludes the paper.

\section{Related Works}
\label{sec:survey}
This section reviews the various approaches to spectrum resource allocation in wireless networks. First, a brief survey of quasi-centralized allocation schemes based on leader election is provided. Then, the fully distributed approaches based on random access protocols are reviewed. Finally, the novel contributions of our work are highlighted.

\subsection{Quasi-Centralized Algorithms}
A naive solution for the spectrum allocation problem in ad-hoc networks is through coordination based on shared CSI between nodes, which reduces the problem of distributed allocation to its centralized form at the cost of excessive signaling overhead. { In the framework of MAB, the centralized approach can be found in~\cite{tibrewal2019multiplayer, boursier2019practical, alatur2020multi, bar2019individual}. In~\cite{tibrewal2019multiplayer}, quantized CSI broadcasting was proposed to enable the implementation of the Hungarian algorithm, since the full knowledge of the CSI among nodes is accessible. A similar algorithm is proposed in~\cite{boursier2019practical} under a hierarchical framework, where a single elected leader node collects the quantized local CSI and dictates the operation of the followers through unicasting. Comparatively, in~\cite{alatur2020multi} the mutually colliding nodes collectively form a ``metaplayer'' and then apply the EXP3 algorithm~\cite{auer2002nonstochastic} to minimize the allocation regret through repeated plays. Also adopting EXP3, information sharing is restricted to the disjoint sub-groups of nodes in~\cite{bar2019individual}, and the mixed strategies of the nodes in a sub-group are computed and dictated by a leading node.
}


\subsection{Distributed Allocation Schemes based on Random Access}
\label{sec_distributed_allocation_reivew}
{
While in theory, the centralized MAB-based algorithms achieve  logarithmic (pseudo) regret, in practice, signaling among distributed nodes for explicit coordination incurs significant overhead. As a result, these methods do not scale well in the case of overly dense or large-scale ad-hoc networks. Furthermore, spectrum coordination may not always be practical since the individual nodes may not be willing to share their local CSI. For these reasons, it is more desirable to develop a fully distributed solution for efficient spectrum allocation without explicit messaging.

The fully distributed, contention-based methods are developed on top of random access protocols such as ALOHA and CSMA/CA. Unfortunately, traditional random-contention ALOHA and CSMA achieve low spectrum efficiency in most practical scenarios, and their performance is especially bad in congested networks~\cite{rom2012multiple}. Hence, decentralized approaches have been put forward to ensure convergence to an optimal orthogonal allocation without explicit information sharing. Examples of these approaches built on ALOHA can be found in~\cite{bistritz2020GOT, wang2020decentralized, bistritz2020my}. Similarly, CSMA-based solutions have been suggested; e.g., in~\cite{leshem2011multichannel, naparstek2013fully,nayyar2016regret, avner2019multi,zafaruddin2019distributed, darak2019multi}. Some studies in this framework aim to achieve optimal resource (e.g., OFDMA resource blocks) allocations, such that the sum of the Quality of Service (QoS) of the users is maximized~\cite{bistritz2020GOT, wang2020decentralized,  nayyar2016regret,  naparstek2013fully, zafaruddin2019distributed}. Comparatively, others aim to achieve stable sub-optimal allocations, with which no user can unilaterally improve its own QoS~\cite{leshem2011multichannel, darak2019multi, avner2019multi}. Recently, new approaches emphasizing fairness among the non-cooperative nodes have also been proposed; e.g., aiming to achieve max-min fairness~\cite{bistritz2020my, bistritz2021one, leshem2023optimal}.
}

\subsubsection{ALOHA-based methods}
{
The distributed spectrum allocation problem can be modeled as a Multi-Agent MAB (MA-MAB) problem, especially when no carrier-sensing capability is assumed for the wireless node, but collisions can be observed. Furthermore, for a highly dynamic radio environment, when the evolution of channel states is independent of node actions, a restless MAB model can be employed by assuming the evolution of each channel to follow an independent Markov model with an unknown transition map~\cite{liu2012learning}.  In this case, to ensure logarithmic regret, extra assumptions regarding the environment or the protocol, such as homogeneous channels for all the users~\cite{6200864, liu2012learning} and pre-agreed round-robin arm-play~\cite{liu2012learning} are typically needed.

A further generalization of the MA-MAB framework can be obtained by emphasizing the impact of link/user strategies on the state change of the radio environment (e.g., from unoccupied to colliding), which leads to a competitive model of Multi-Agent MDP (MA-MDP)~\cite{7429691} or to potential games \cite{cohen2013game, cohen2015distributed}. In~\cite{naparstek2018deep, kwasinski2020, yu2019deep}, an empirical solution framework based on multi-agent independent Deep Reinforcement Learning (DRL) is adopted. However, these techniques do not have a theoretical guarantee for convergence and suffer linear regret, since they are inherently sub-optimal.
}
In~\cite{bistritz2020GOT}, a cooperative framework exploiting collisions is proposed to obtain logarithmic regret in the multi-agent setup without explicit communication.   In~\cite{wang2020decentralized}, the MA-MAB game is extended to contextual bandits, which addresses the arbitrary evolution of radio environment states for IoT devices operating on unlicensed channels. The same approach using bandit-based games can also be found in~\cite{9466123} for ALOHA-based power-domain non-orthogonal resource allocation.

\subsubsection{Carrier sensing-based methods}
ALOHA-based methods and their variants rely solely on the local link feedback upon collisions. Comparatively, enabling the on-device capability of \emph{listen-before-talk}; e.g., with CSMA, may significantly improve the time efficiency of the allocation schemes with a built-in collision avoidance mechanism. In~\cite{leshem2011multichannel}, the Gale-Shapley algorithm~\cite{galeshapley1962} based on opportunistic CSMA/CA~\cite{zhao2005opportunistic} is employed to distributedly find a stable one-to-one matching between users and static channels in a single contention window with a deterministic number of iterations. In~\cite{hossler2020stable, holfeld2016stable}, this approach is extended to the cases of many-to-one matching and many-to-many matching, respectively, where more than one channel are assigned either to a single user or to multiple users simultaneously.

{
For time-varying channels, swapping-based algorithms, which also operate in the framework of MA-MAB, have been developed for distributed learning of stable allocations~\cite{leshem2011multichannel, darak2019multi,avner2019multi}. By enabling carrier sensing, these algorithms are designed to address the situation of users frequently leaving and joining the network. Thus, instead of optimal allocation, they aim to ensure stable orthogonal allocations with high probability over time. For this reason, regret is linear with respect to the optimal allocation.
}

To obtain the socially optimal spectrum allocation, opportunistic CSMA is employed as the building block for implementing the well-known distributed auction algorithm~\cite{bertsekas1979distributed} in ad-hoc wireless networks. In~\cite{zafaruddin2019distributed}, a distributed channel auction is performed based on the locally estimated QoS over each channel by the devices, for which the accuracy of their local estimation improves with repeated plays in an MA-MAB. Without an auction arbitrator, the QoS-dependent back-off in CSMA plays a key role in distributedly indexing the bids of different devices. This work can be considered an extension of the distributed auction from the scenario of networks with static channels (e.g.,~\cite{naparstek2013fully, zappone2016distributed}) to the scenario with dynamic channels. Note that most studies using a distributed auction for strategy learning in an MA-MAB framework need explicit signaling through either unicasting~\cite{nayyar2016regret} or broadcasting~\cite{8440092} to index the bids offered by the decentralized agents.

\subsection{Contribution and Novelty of Our Work}
Apart from the dense network scenario that is not dealt with in the literature, the main contributions of our access-strategy learning framework include:
\begin{enumerate}
    \item Our proposed protocol allows distributed optimal time-frequency allocation during strategy exploitation where transmission is orthogonal.
    This scheme guarantees the logarithmic regret in time (i.e., $O(\ln T)$) when the time consumed by strategy exploitation grows exponentially. This is achieved with a bounded-size contention window and finite bid resolution during the distributed auction for optimal resource allocation.
    \item We provide complete framing and state-machine design for the main procedures of the proposed allocation protocol. To accelerate the distributed auction, we introduce the technique of $\varepsilon$-scaling for the bid proposal to reduce the number of iterations needed for auction convergence. 
    \item To design an implementation-friendly protocol, we modify our proposed scheme by adopting a fixed frame size for each epoch. Our analysis shows a bounded strategy-exploration overhead for this protocol. Moreover, the overhead can be arbitrarily low, and the probability of obtaining sub-optimal allocation decreases exponentially with time. This design improves the network scalability since the links can quickly adapt to network changes.
\end{enumerate}

\section{System Model}
\label{sec:system}
\begin{figure}[t!]
  \centering
  \includegraphics[width = 0.8\columnwidth]{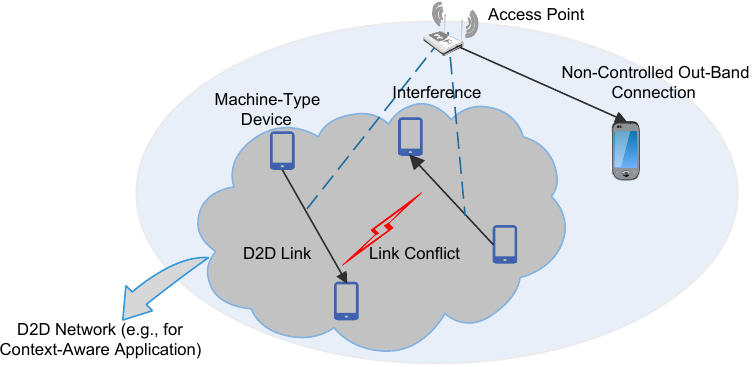}
  \caption{An example of the D2D ad-hoc network.}
  \label{fig:Network_Overview}
\end{figure}

\begin{figure}[t!]
  \centering
  \includegraphics[width = 0.9\columnwidth]{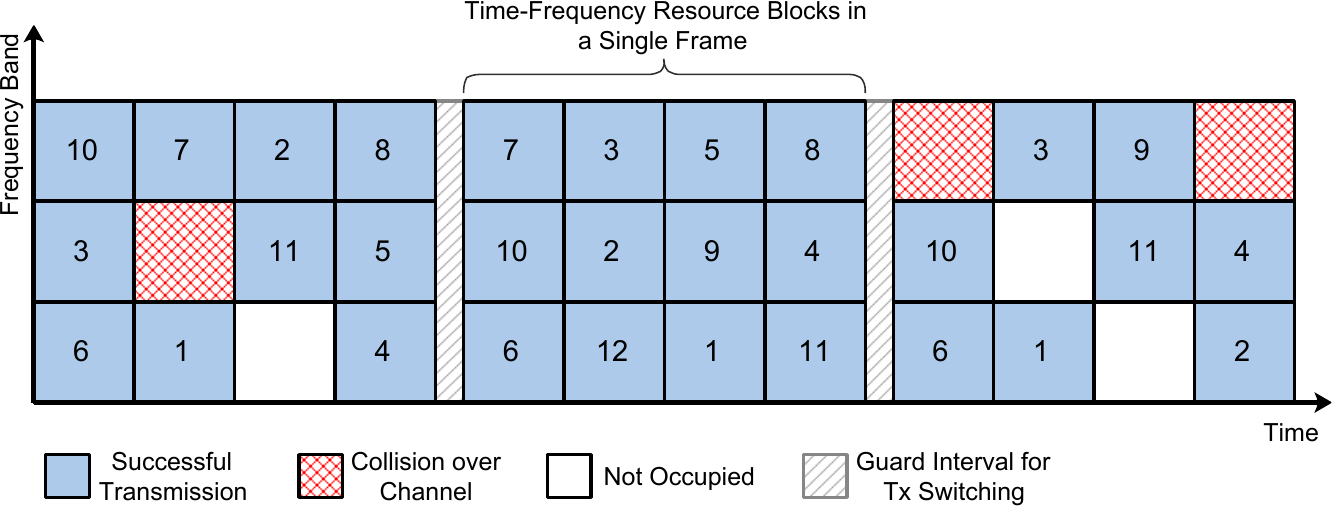}
  \caption{Example of contention-based allocation of $3\times4$ time-frequency resource blocks. The numbers over each resource block represent the link IDs.}
  \label{fig:OFDMA}
\end{figure}

{
Consider a D2D ad-hoc network as in Figure~\ref{fig:Network_Overview}, where $N$ synchronized D2D links operate over an (unlicensed) joint frequency band of $K$ frequency channels. We assume that the wireless nodes are densely deployed and $N>K$. Therefore, even when the D2D links operate with low traffic loads, time division is still necessary. We assume that each frame is composed of $M=\lceil N/K\rceil$ time slots. The network utilizes a CSMA-like protocol during the allocation-strategy learning phase and aims to provide an optimal orthogonal allocation during the data transmission phase. Let $\cK=\{1,\ldots,K\}$ denote the set of channels, and $\cM=\{1,\ldots, M\}$ denote the set of time slots in a single frame. This leads to an OFDMA-like configuration as depicted in Figure~\ref{fig:OFDMA}.

We assume that each transmitter-receiver pair can listen to and transmit over at most one time-frequency resource block during a single frame due to on-device battery and RF-chain constraints. At each frame, a link $n$ acts by selecting a time-frequency pair $a_n=(k, m)$. Hence the action space of each link is $\cA=\cK\times\cM$, leading to a joint action profile of the $N$ links  $\mathbf{a} \triangleq [a_1,..., a_N]^{\textrm{T}}$. We also assume that during each time slot within a frame, the gain of each link over channel $k\in\cK$ remains constant. However, due to random interference, the links are assumed to experience heterogeneous and unknown stationary channel evolution over different resource blocks across different time slots.}

Practically, a wireless node chooses a fixed strategy from a relatively small set of coding and modulation schemes during its operation. This implies that the achievable QoS level belongs to a small set $\{\underline{q}, \ldots, \overline{q}\}$. Without loss of generality, we assume that these QoS levels are all integer multiples of a basic service level $\gD_{\min}$. The instantaneous QoS is estimated by the receiver and fed back to the transmitter. Let $\cN$ denote the set of links, and $q^t_n$ denote the instantaneous QoS level (e.g., throughput) of link $n\in\cN$ at frame $t$. By our assumption, we have $\forall n\in\cN, 0\le q^t_n\le \overline{q}$. Furthermore, let $\Delta$ be the resolution of the measured QoS relative to $\overline{q}$, and we have
\begin{equation}
  \label{def:delta}
  \gD \triangleq \gD_{\min} / \overline{q}.
\end{equation}

{ For a dense network, it is reasonable to assume that whenever two links use the same resource block simultaneously, they will block each other and fail to transmit in the corresponding time slot. We define the instantaneous utility of link $n$ with action $a_n$ at time frame $t$ as its achieved QoS level:
\begin{equation}
\label{eq:def:utility}
    u^t_n \left(\mathbf{a}\right) \triangleq \left\{
    \begin{array}{ll}
      q^t_n(a_n), & \textrm{if } \forall m\in \cN/\{n\}, a_m\ne a_n,\\
      0, & \textrm{otherwise},
    \end{array}\right.
\end{equation}}
By our assumption, the QoS of stochastic channels, $q^t_n(a_n)$ are time-varying for a given action $a_n$. We are interested in optimizing the sum of the long-term averages of the utilities of all the links. Since the statistical properties of the stationary resource blocks are not known, each link has to learn the expected values of its QoS through repeatedly sampling the QoS feedbacks over different channels. By our assumption, $q^t_n(a_n)$ is i.i.d. in time and sampled from $[0,\overline{q}]$ following an unknown distribution.

We aim to find an action profile $\va^*$ that optimizes the expected social utilities of the D2D network during its operation:
\begin{align}
  \label{eq_objective}
    \va^*=\arg\max_{\va} W(\va)=\Ep\limits_{\forall n: q_n}\left\{\sum_{n=1}^N u_n(\va)\right\},
\end{align}
where $u_n(\va)$ is determined by \eqref{eq:def:utility}, and $q_n$ is the instantaneous QoS of link $n$ over the selected resource block $a$ without any collision. From \eqref{eq_objective}, let $W^*\triangleq W(\va^*)$ denote the optimal expected average social utility. This serves to define the social regret of the links caused by adopting an arbitrary sequence of joint actions $\{\va_t\}_{t=1}^T$ as follows:
\begin{align}
  \label{eq_regret}
  R \triangleq TW^* - \displaystyle\Ep \left\{ \sum_{t=1}^{T} \sum_{n=1}^N u_n(\va_t)  \right\},
\end{align}
where the expectation is taken with respect to the randomness of the sampled utilities as in~\eqref{eq_objective}. Thus, the goal of our study can be transformed from the centralized optimization problem in~\eqref{eq_objective} into finding a distributed solution to the problem of minimizing the regret as given in~\eqref{eq_regret}.


\section{Protocol Structure for MAB-Based Distributed Resource Allocation}
\label{sec:protocol}
To distributedly minimize the regret in~\eqref{eq_regret}, the active links need to complete two major tasks: (a) to obtain the accurate estimates of their expected QoS levels over each resource block, and (b) to determine the socially optimal strategies through CSMA-based contentions. The former task aims to accumulate QoS samples over each resource block from the successful transmissions, so an accurate index of the resource blocks can be obtained. The latter requires the links to learn the optimal orthogonal strategies as fast as possible without any information exchange between the ad-hoc links. To achieve these goals, we introduce an MA-MAB framework to address the tradeoff between channel-quality index exploration and myopic orthogonal-strategy exploitation based on the estimated QoS levels.

\begin{figure}[!t]
  \centering
  \includegraphics[width = 0.90\columnwidth]{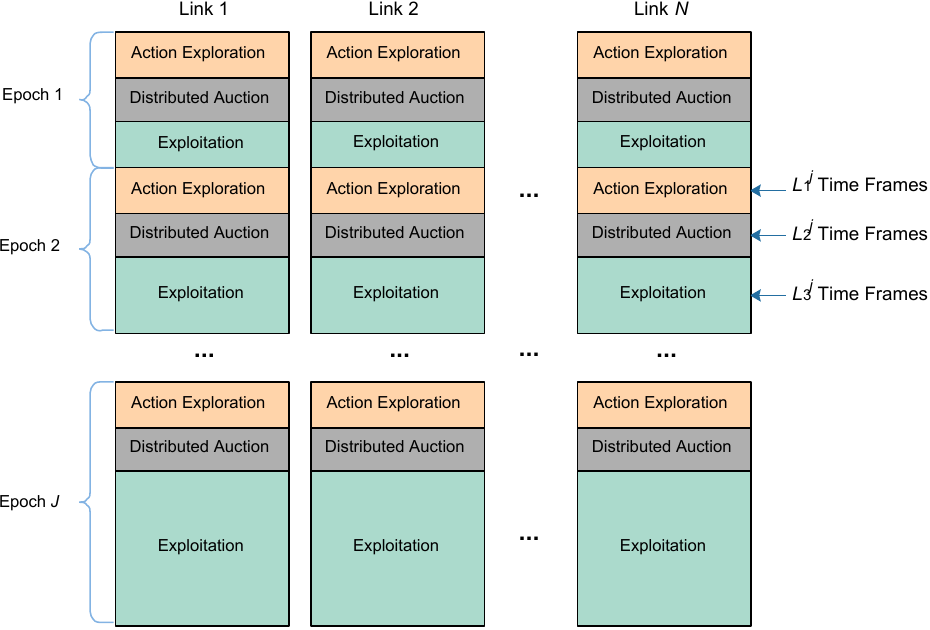}
  \caption{Synchronized link actions in repeated epochs of three phases, where the number of time frames in each phase may vary.}
  \label{fig:new_phases}
\end{figure}

We consider a scenario where the total network-operation period, $T$ in~\eqref{eq_regret}, is not known in advance. We run decentralized exploration and exploitation of the link strategies in repeated epochs as in~\cite{zafaruddin2019distributed}, by treating each link as an agent and each resource block as an arm in the MA-MAB. The ad-hoc links are synchronized across epochs, and each epoch is divided into three phases: (a) exploration, (b) coordination for optimal resource allocation through distributed auction and (c) exploitation. The exploration phase is designed for each transmitter to independently learn the QoS over different resource blocks, while the coordination phase is used to obtain an optimal allocation with respect to the link-QoS information learned in the exploration phase. { During the exploitation phase, each link selects the action it obtained from distributed auction phase and transmits data over this time-frequency resource.} Let $j$ ($j\!=\!1,2,\ldots,J$) denote the epoch index, and $L^j_1$, $L^j_2$ and $L^j_3$ denote the (possibly variable) number of time frames in the three phases of epoch $j$, respectively. We depict the organization of the time frames in an epoch in Figure~\ref{fig:new_phases}, which maps into a three-phase algorithm for a single link in Algorithm~\ref{alg:1}. In what follows, we will explain the detailed design of the link schemes for action selection in Algorithm~\ref{alg:1}.

\begin{algorithm}[t]
  \begin{small}
  \caption{Epoch-based learning for resource-block selection by link $n$}
\begin{algorithmic}[1]
  \STATE \textbf{Initialize:} For all $a\in\cA$ set the counter of action visitation as $v_{n,a}=0$ and the accumulated reward for action $a$ as $r_{n,a}=0$.
  \FOR{epoch $j=1$ \textbf{to} $J$}
  \STATEx \textit{Exploration phase:}
    \STATE Update the vectors of the visitation counts $\vv_n$ and the accumulated rewards $\vr_n$, and the matrix of the estimated QoS over each channel $\hat{\mathbf{Q}}_n=[\hat{q}_{n,a}]_{a\in\cA}$:
    \begin{equation}
      \label{eq_exploration_output}
      \left[\hat{\mathbf{Q}}_n, \vv_n, \vr_n \right] \leftarrow \textbf{Exploration}(\cK, \cM, \vv_n, \vr_n).
    \end{equation}
    \STATEx \textit{Coordination (distributed auction) phase:}
    \STATE Update the local action $a_n$:
    \begin{equation}
      \label{eq_auction_output}
      a_n  \leftarrow  \textbf{Auction}\left(\hat{\mathbf{Q}}_n\right).
    \end{equation}
    \STATEx \textit{Exploitation phase:}
    \FOR {time slot $t=1$ \textbf{to} $L_j^3$}
      \STATE {  Fix the local action to $a_n$ as obtained in (\ref{eq_auction_output}) for data transmission.}
    \ENDFOR
  \ENDFOR
\end{algorithmic}
\label{alg:1}
\end{small}
\end{algorithm}
\subsection{Exploration phase}
\label{subsec:exploration_phase_protocol}
During the exploration phase, a transmitter collects data regarding the mean QoS on each resource block through repeated sampling. This is done by choosing a pair of the transmitting channels and time slots uniformly at random\footnote{The coordination of random frequency-time selection between the end nodes of an ad-hoc link can be achieved by adopting a common pseudorandom generator using the same random seed on both ends.}. For simplicity of implementation, we consider that the links adopt an ALOHA-like mechanism for collision detection in this phase. More specifically, the transmitter attempts to transmit over a randomly selected resource block by sending a series of pilot signals. When no collision occurs over the target resource block, and the pilot signal is properly received, an ACK message containing the estimated QoS is sent back by the receiver to the transmitter. Otherwise, the transmitter knows that a collision with some other links occurred. This operation is formalized by Lines 4-11 in Algorithm~\ref{alg:2}.
\begin{algorithm}[t]
  \begin{small}
  \caption{$\textbf{Exploration}(\cK, \cM, \vv_n, \vr_n)$: The exploration phase for link $n$ in epoch $j$}
  \begin{algorithmic}[1]
  \STATE \textbf{Input:} $\cK,\cM, \vv_n,\vr_n$. \hfill\COMMENT{time-frequency sets, the vectors of \newline
      \hspace*{8em} visitation counts and accumulated rewards}
  \STATE \textbf{Output:} $\hat{\mathbf{Q}}_n, \vv_n,\vr_n$. \hfill\COMMENT{the matrix of local QoS estimates and  \newline
  \hspace*{2em} the vectors of visitation counts and accumulated rewards}
  \STATE \textbf{Initialize:} For all $a$, set the sampling rule for the dithering value as $d_{n,a} \sim U \left(\left[-d_{\max}, d_{\max}\right]\right)$, where $d_{\max}=\gD_{\min}/8N$.
  \FOR{time slot $t=1$ \textbf{to} $L_1^j$}
      \State {Transmitter $n$ chooses an action $a=(k\in\cK,m\in\cM)$ uniformly at random and sends pilots over resource block $a$.}
      \IF {pilots are received by receiver $n$}
        \STATE Receiver $n$ evaluates the instantaneous QoS as $\hat{u}^t_{n}(a)$ and sends an ACK message containing $\hat{u}^t_{n}(a)$.
      \ENDIF
      \IF {ACK message is received}
           \STATE  Transmitter $n$ updates the visitation count $v_n$ and the accumulated reward $r_{n,a}$ in $\vv_n$ and $\vr_n$, respectively:
           \begin{equation}
             \label{eq:estimation_step_update}
             \left\{
             \begin{array}{ll}
                v_{n,a}\leftarrow v_{n,a} +1, \\
                r_{n,a}\leftarrow r_{n,a} + \hat{u}^t_{n}(a),
             \end{array}\right.
           \end{equation}
      \ENDIF
  \ENDFOR
\STATE { $\forall a: v_{n,a}\ne 0$: transmitter $n$ updates the estimated QoS of resource block $a$ in $\hat{\mathbf{Q}}_n$ with a random dithering value $d_{n,a}$:}
  \begin{equation}
    \label{eq_update_estimate}
    \hat{q}_{n,a}\leftarrow r_{n,a}/v_{n,a} + d_{n,a}.
  \end{equation}
\end{algorithmic}
\label{alg:2}
\end{small}
\end{algorithm}

Subsequently, if an ACK is received for the attempted action, transmitter $n$ updates its local estimation of the QoS over the selected resource block $a=(k,m)$ according to~\eqref{eq:estimation_step_update} and (\ref{eq_update_estimate}). Note that to prepare the local QoS estimates for distributed auction in the coordination phase, transmitter $n$ adds a random dithering value $d_{n,a}$ to the local estimation in (\ref{eq_update_estimate}). The dithering value helps to break potential ties in resource-block indexing both locally and among different links. In Section~\ref{sec:regret}, Lemma~\ref{lem:sufficient_utility_estimation_precision} will show that by admitting a proper dithering range $d_{\max}$, the action obtained using the dithered QoS estimates aligns with the optimal allocation strategies. The common dithering range across the network can be initialized at the beginning of random access with very low overhead. Since this is a standard procedure in ad-hoc networks, we assume that this information is known at the beginning of the network operation.

\subsection{Coordination for Orthogonal Resource Allocation in Distributed Auction}
\label{subsec:auction_phase_protocol}
We now describe the distributed auction phase. By the end of the exploration phase, each link $n\in\cN$ has an estimated QoS for every resource block; i.e., $\forall a\in\cA=\cK\times\cM$. Let $\hat{\mathbf{Q}}_n=[\hat{q}_{n,a}]_{a\in\cA}$ denote the matrix of the estimated QoS by link $n$ in Algorithm~\ref{alg:2}. To obtain an optimal allocation without exchanging the action information among the links, we apply distributed auction~\cite{naparstek2013fully} and derive an algorithm to tackle the practical issues of bid quantization with finite accuracy (i.e., represented by the discrete back-off interval).

\begin{figure}[t]
    \centering
    \includegraphics[width=0.4\textwidth]{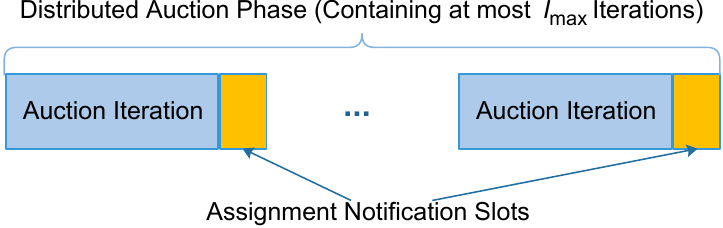}
    \caption{ The structure of the auction phase. Unassigned users transmit on the common channel (channel 0) in the assignment notification slot. The maximum number of iterations $I_{\max}$ is determined later in Lemma~\ref{lem:finite_time_convergence}.}
     \label{fig:Auction_Phase_structure}
\end{figure}

\begin{figure}[t]
  \centering
    \includegraphics[width=0.3\textwidth]{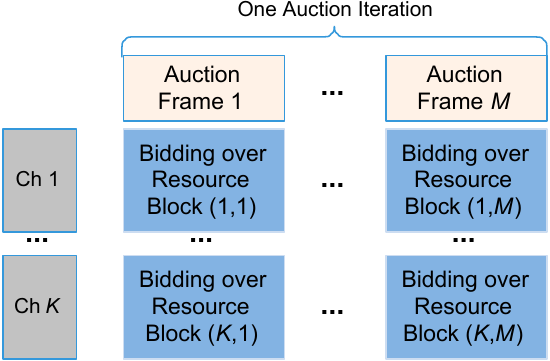}
        \caption{ Each auction iteration is composed of $M$ auction frames. The nodes bid over resource $a=(k,m)$ in channel $k\in\cK$ and frame $m\in\cM$.}
\label{fig:bidding_over_different_channels_simultaneously}
\end{figure}

\begin{figure}[t]
  \centering
    \includegraphics[width=0.35\textwidth]{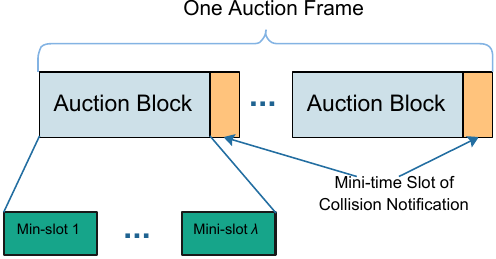}
        \caption{ The structure of auction frames and auction blocks.}
\label{fig:Auction_Frame}
\end{figure}

{  We consider that $\forall n\in\cN$, link $n$ maintains two binary states for the resource assignment process, $f_n\in\{0, 1\}$ and $f_n^g\in\{0, 1\}$. $f_n$ is an assignment flag indicating whether link $n$ itself is granted access to some resource block ($f_n=1$) or not ($f_n=0$). $f^g_n$ is a locally-kept global state indicating whether the resource assignment for all the links is completed ($f^g_n=1$) or not ($f^g_n=0$).

The auction phase is executed in parallel by all nodes in the network. The structure of the distributed auction protocol is now described in Algorithm~\ref{alg:3}.
Each auction phase is composed of auction iterations as described in Figure~\ref{fig:Auction_Phase_structure}. After each iteration, each assigned node listens during a single assignment notification slot over a common channel (channel 0). Each unassigned node transmits an ACK notification over channel 0 during this time slot.  If a transmission is heard the assignment flag is off ($f^g_n=0$). If no message is transmitted, all the nodes move to the exploitation phase. Each auction iteration is composed of frames in time and frequency as described in Figure~\ref{fig:bidding_over_different_channels_simultaneously}. A single frame implements the bidding over all time frequency channels in parallel. Furthermore, the auction frames are composed of auction blocks (see Figure~\ref{fig:Auction_Frame}), which implement the bidding over a single time-frequency block. One auction block is composed of multiple contention windows of length $\log_2\gb$ followed by a collision indicator.}

During the auction, the (virtual) net profit/gain of link $n$'s successful transmission over resource block $a=(k,m)$ can be estimated based on the local valuation matrix of the resource blocks $\hat{\mathbf{Q}}_n$:
\begin{equation}
\label{eq:profit_def}
    g_{n,a} = \hat{q}_{n,a} - b_{n,a},
\end{equation}
where $b_{n,a}$ represents the virtual price that link $n$ is willing to pay for the assignment of resource $a=(k,m)$. Let $\mathbf{B}_n=[b_{n,a}]_{a\in\cA}$ denote the matrix of bids placed by link $n$ for the resource blocks.
During the distributed auction formalized in Algorithm~\ref{alg:3}, the operation of a link $n$ (in parallel with the other links) is repeated in three stages: ordering of the locally estimated rewards over each resource block (Lines 6-8), winner determination after the placement of the quantized bids (Lines 9-14) and monitoring of the state of the distributed auction (Lines 15-19). In Section~\ref{sec:regret} we will show that the number of iterations required for the auction phase is upper-bounded by $I_{\max}=\frac{8N^3\overline{q}}{\dm}(1+\frac{1}{8 N})$ (see Lemma~\ref{lem:finite_time_convergence}). Distributed monitoring of the auction state is completed in the dedicated assignment notification slot (see Figure~\ref{fig:Auction_Phase_structure}), which relies on the channel-sensing capability of the wireless nodes for determining whether the auction is completed.
\begin{algorithm}[!t]
  \begin{small}
    \caption{$\textbf{Auction}\left(\hat{\mathbf{Q}}_n\right)$: Bidding in the distributed auction by link $n$ in epoch $j$}
    \begin{algorithmic}[1]
      \STATE \textbf{Input:} $\hat{\mathbf{Q}}_n$. \hfill\COMMENT{the local matrix of estimated QoS}
      \STATE \textbf{Output:} $a_n$. \hfill\COMMENT{the selected resource block}
      \STATE \textbf{Initialize:} Set the initial step of bid increment as  $\varepsilon = \dm/4$, and the minimum step of bid increment as $\varepsilon^* = \dm/8N$. Set the initial bid matrix as $\mathbf{B}_n=\vzero$. Set the local indicators of assignment and auction completion to be $f_n=0$ and $f^g_n=0$, respectively. Choose a bid-step scaling factor $0<\zeta<1$.
      \WHILE{$f^g_n=0$} \hfill\COMMENT{auction is not completed}
          \IF{$f_n=0$}\hfill\COMMENT{assignment is not determined}
            \STATE {Compute the profit matrix over all the candidate resource blocks according to (\ref{eq:profit_def}):
            \begin{equation}
              \label{eq_profit_matrix}
              \mathbf{G}_n = [g_{n,a}=\hat{q}_{n,a} - b_{n,a}]_{a\in\cA},
            \end{equation}}
            \STATE Select the best action $\tilde{a}_n$ and its corresponding profit ${g}_{n,\tilde{a}_n}$:
            \begin{equation}
              \label{eq:most_profitable_resource_def}
                \tilde{a}_n = \arg\max_{a\in\cA} g_{n,a},
            \end{equation}
            \STATE Find the second-best action $a'_n$, such that
            \begin{equation}
              \label{eq:second_most_profitable_resource_def}
                {a}'_n = \arg\max_{a\in\cA/\{\tilde{a}_n\}} g_{n,a},
            \end{equation}
            and its corresponding profit $g_{n,a'_n}$.
            \STATE Update the bid on resource block $\tilde{a}_n$ in $\mathbf{B}_n$ as follows:
            \begin{equation}
            \label{eq:new_bid}
                b_{n,\tilde{a}_n}  \leftarrow b_{n,\tilde{a}_n} + \left( \epsilon + {g}_{n,\tilde{a}_n} - g_{n,a'_n}\right).
            \end{equation}
          \ELSE
            \STATE Stay with the same action: $\tilde{a}_n\leftarrow a_n$.
          \ENDIF
          \STATE Update the bid step $\varepsilon$:
          \begin{equation}
          \label{eq:epsilon_scaling}
              \varepsilon \leftarrow \max \left\{\varepsilon^*, \zeta\varepsilon \right\}.
          \end{equation}
          \STATE Place the (quantized) bid on the candidate action $\tilde{a}_n$:
          \begin{equation}
            \label{eq_place_bid}
            [a_n, f_n] \leftarrow  \textbf{Bid}\left(\tilde{a}_n, \mathbf{B}_{n} \right)
          \end{equation}
          following Algorithm~\ref{alg:5}.
          \IF{$f_n=0$}
            \STATE Transmit in the assignment notification slot (blocking).
          \ELSE
            \STATE Listen to the assignment notification slot. Set $f^g_n=1$ when no transmission is overheard.
          \ENDIF
    \ENDWHILE
    \end{algorithmic}
  \label{alg:3}
  \end{small}
\end{algorithm}

\begin{figure}[t]
\centering
\includegraphics[width = 0.8\columnwidth]{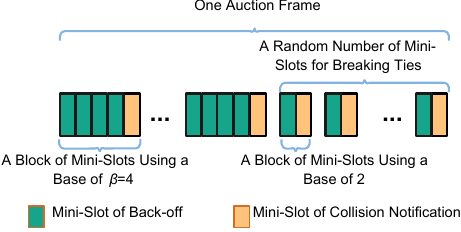}
\caption{ The detailed structure of contention window in one auction frame. A contention window contains a fixed number and a random number of sub-slots for back-offs to determine the order of bids from different links.}
\label{fig:Auction_Phase}
\end{figure}

{ 
We note again that the distributed auction phase runs on the time scale of auction iterations. From the network perspective, one iteration in an auction phase over a given channel is further divided into $M$ auction frames. For each link, bidding over a resource block $a=(k,m)$ is performed over channel $k$ in the $m$-th auction frame (see Figure~\ref{fig:bidding_over_different_channels_simultaneously}). To resolve the winner determination problem over $a=(k,m)$, we introduce a CSMA contention window for each auction frame. A contention window is composed of two parts for transmission back-off (see Figure~\ref{fig:Auction_Phase}): a fixed number of long auction blocks and a random number of short auction blocks.  This division is aimed to keep the number of contention slots in each block small ($\gb$) so the bid is done sequentially, intially using the first $\gb$ most significant digits and if there is not single winner bidding continue in the next auction blocks. By the end of the fixed number of blocks it is very likely that we have a winner. However, in case there is no winner a standard binary collision resolution takes place.
Each block contains several mini-slots and is used for each link to back-off based on its bid over the time-frequency resource of interest. During a contention window, transmitter $n$ determines its back-off period as $\tau_{n,a}=f(b_{n,a})$ based on a predetermined, monotonically decreasing function $f(\cdot)$ and its local bid $b_{n,a}$ for action $a$. The selection of resource block $a$ by transmitter $n$ is allowed if and only if no other transmission is overheard on $a$ before $\tau_{n,a}$ expires.
}

\begin{algorithm}[t]
  \begin{small}
  \caption{$\textbf{Bid}\left(\tilde{a}_n, \mathbf{B}_{n}\right)$: CSMA-based winner determination using bid quantization for link $n$}
  \begin{algorithmic}[1]
    \STATE \textbf{Input:} $\tilde{a}_n, \mathbf{B}_{n}$.
    \STATE \textbf{Output:} $a_n, f_n$. \hfill\COMMENT{resource ID and state flag}
    \STATE \textbf{Initialize:} Set the local indicators of assignment to be $f_n=0$. Set the local indicator of winner determination to be $w_n=0$ ($w_n\in\{0, 1\}$). Set the block index to be $b_n=1$.
    \WHILE{$w_n=0$} \hfill\COMMENT{winner is not determined}
        \IF{$f_n=0$}
            \IF{$b_n \le \lambda$}
                \STATE Choose the back-off time length $\gr_{b_n}$ in the $b_n$-th back-off block according to~\eqref{eq:new_quant}, for which the input $\tau_{n, \tilde{a}_n}$ is given as $\tau_{n, \tilde{a}_n}=f(\mathbf{B}_{n,\tilde{a}_n})$ in (\ref{eq_backoff_generator}).
            \ELSE
                \STATE Choose the back-off time length uniformly at random.
            \ENDIF
            \STATE Sense before transmission after the back-off mini-slots.

            \IF{some other transmission is sensed}
                \STATE $f_n\leftarrow 0$, $w_n\leftarrow 1$ and $a_n=0$. \hfill\COMMENT{losing the auction}
            \ENDIF
            \IF{ACK received in the collision notification slot}
                \STATE $f_n\leftarrow 1$, $w_n\leftarrow 1$ and $a_n=\tilde{a}_n$. \hfill\COMMENT{winning the auction}
            \ENDIF
        \ENDIF

        \STATE $b_n\leftarrow b_n+1$.
    \ENDWHILE
  \end{algorithmic}
  \label{alg:5}
  \end{small}
\end{algorithm}

It is worth noting that each link updates its bid on the target resource block discretely with a minimum increment step $\varepsilon$. $\varepsilon$ is updated in each bid iteration according to (\ref{eq:epsilon_scaling}) with a scaling factor $0<\zeta<1$ until a minimum step size $\varepsilon^* \triangleq \dm/8N$ is reached. The $\varepsilon$-scaling can significantly accelerate the convergence of the auction algorithm. Lemma~\ref{cor:optimal_finite_convergence} in Section~\ref{sec:regret} shows that this step size guarantees the convergence to the socially optimal allocation. During the  bid placement procedure, $\textbf{Bid}(\cdot)$ in Algorithm~\ref{alg:5}, transmitter $n$ translates the continuous bid obtained from (\ref{eq:new_bid}) into a discrete value of finite digits. Using the proposed bid-step adaptation scheme and $\varepsilon^*$, the maximum rounds of discrete bids that a link can make, $N_b^*$, is given by
\begin{equation}
\label{eq:maximal_discrete_bid}
    N_b^* = 8 N \overline{q} / \dm.
\end{equation}

Consequently, we adopt the following continuous back-off generation function $\forall n\in\cN$:
\begin{align}
  \label{eq_backoff_generator}
  \gt_{n,a}=f(b_{n,a})=1-b_{n,a}/\overline{q}.
\end{align}
The continuous backoff time $\tau_{n,a}$ ($0<\tau_{n,a}<1$) needs to be quantized, such that it can be identified within a finite number of mini-slots. We quantize $\tau_{n,a}$ using a base-$\beta$ representation as follows:
\begin{align}
\label{eq:new_quant}
    [\gt_{n,a}]_{\gb, \gl} &= \gr_1\times\beta^{-1}+\gr_2\times\beta^{-2}+\ldots+\gr_\gl\times\beta^{-\gl} \nonumber \\
    & \triangleq 0.\gr_1 \gr_2\ldots \gr_\gl,
\end{align}
where each digit $\gr_i$ ($1\le i\le\gl$) can be mapped into a group of $\beta$ mini-slots\footnote{Our experiments shows that $\beta=4$ leads to very good performance.}, and $\gl$ is defined as the resolution (in the number of retained quantization digits) of the bids in the auction algorithm with respect to $\varepsilon^*$:
\begin{equation}
    \label{eq_number_of_digits}
    \lambda = \log_\beta N_b^*=\log_\beta\left(8 N \overline{q} / \dm\right).
\end{equation}

{ 
Thus, the contention window of an auction frame contains at most $\gl$ fixed blocks, each of which contains $\gb$ grouped mini-slots for back-off and one collision notification mini-slot for winner determination (see Figure~\ref{fig:Auction_Phase}) over its corresponding digit in (\ref{eq:new_quant}). To break the potential ties of the highest bids due to the accuracy loss (truncated at an accuracy of $\overline{q}/b^*$) caused by quantization with (\ref{eq_number_of_digits}), the $\gl$ blocks of mini-slots can be followed by a random number of collision resolution blocks. Each collision resolution block contains one back-off mini-slot and one collision notification mini-slot. We note that after $\gl$ auction blocks, if there still remain more than one candidate links, there is no preference to the remaining links due to precision truncation. Then, the collision resolution blocks are needed. With random collision resolution, we randomly select one of the remaining links as the auction winner. Lemma~\ref{lem:blocks} in Section~\ref{sec:regret} will show that the number of auction blocks in this random back-off part is upper bounded by $\log(N)$. Note that in practice we do not expect more than $N/K$ users on avergae at each auction frames which further reduces the number of collisions.

For a single frame in the auction, the CSMA-based winner determination mechanism is described in Algorithm~\ref{alg:5}. As indicated by Lines 4-19, in a collision resolution block $b_n$ ($1\le b_n\le\gl$), link $n$ chooses a back-off of $\gr_{b_n}$, mapped to the mini-slots before transmitting, where $\gr_{b_n}$ corresponds to the quantized digit and is determined by~\eqref{eq:new_quant}. If any other link finishes its back-off before link $n$, link $n$ loses its bidding on the selected resource block $\tilde{a}_n$ and terminates the back-off process with $f_n=0$ (see Line 13). After a back-off block of $\gb$ mini-slots, a mini-slot of collision notification is dedicated to the receivers, which fail to decode the intended message, to transmit a NACK message. If no transmission occurs in this mini-slot, the single link winning the back-off contention changes its state to $f_n=1$, indicating the completion of its bidding over the resource in a single auction iteration. Otherwise, the back-off contention continues to the next group of minislots with each remaining transmitter using the next digit as the back-off if $b_n<\gl$, or on a random number of blocks if the accuracy of quantization needed for distinguishing the maximum bids exceeds the allowable level $\gl$ (see also Lemma~\ref{lem:optimal_convergence} in Section~\ref{sec:regret}). Accompanying Algorithm~\ref{alg:5}, we also provide a schematic description of the collision-resolution scheme in Figure~\ref{fig:auction_block_flow}.}
\begin{figure}[!t]
  \centering
\begin{tikzpicture}[node distance=1.5cm, align=center]
  \tikzstyle{startstop} = [rectangle, rounded corners, minimum width=1cm, minimum height=0.5cm,text centered, draw=black, fill=pink!30]
  \tikzstyle{io} = [trapezium, trapezium left angle=70, trapezium right angle=110, minimum width=1cm, minimum height=0.5cm, text centered, draw=black, fill=gray!30]
  \tikzstyle{process} = [rectangle, minimum width=1cm, minimum height=0.5cm, text centered, draw=black, fill=green!10]
  \tikzstyle{decision} = [diamond, minimum width=0.5cm, minimum height=0.5cm, text centered, draw=black, fill=blue!10]

  \tikzstyle{arrow} = [thick,->,>=stealth]
  \node (initialization) [startstop, text width = 1.5cm,] {\tiny\linespread{0.8} Set local state $f_n=0$, $b_n=1$\par};
  \node (pro_backoff) [process, text width = 2.5cm, right of= initialization, xshift=1.0cm] {\tiny Back-off for $\gr_{b_n}$ or a random number of mini-slots};
  \node (dec_trans_dec) [decision, aspect=2.0, right of= pro_backoff, xshift=1.2cm] {\tiny TX detected \par};

  \node (output1) [startstop, text width = 1.5cm, below of=dec_trans_dec, yshift=-0.1cm, xshift=1.6cm] {\tiny Terminate with $f_n=0$, $a_n=0$\par};
  \node (dec_ack) [decision, aspect=2.0, left of= output1, xshift=-0.5cm] {\tiny ACK received \par};
  \node (pro_continue) [process, text width = 1.5cm, below of= pro_backoff, yshift=0.5cm, xshift=0.0cm] {\tiny  ${b_n}\leftarrow {b_n}+1$};
  \node (output2) [startstop, text width = 1.5cm, below of=initialization, yshift=-0.1cm] {\tiny Terminate with $f_n=1$, $a_n=\tilde{a}_n$\par};

  \draw [arrow] (initialization) -- (pro_backoff);
  \draw [arrow] (pro_backoff) -- (dec_trans_dec);

  \draw [arrow] (dec_trans_dec)($(dec_trans_dec.east)$) -- node [anchor=west]{\tiny Yes} (output1);
  \draw [arrow] (dec_trans_dec)($(dec_trans_dec.south)$) -- node [anchor=west]{\tiny No} (dec_ack);

  \draw [arrow] (dec_ack)($(dec_ack.west)$) -- node [anchor=north]{\tiny Yes} (output2);
  \draw [arrow] (dec_ack) -- node [anchor=south]{\tiny No} (pro_continue);
  \draw [arrow] (pro_continue) -- (pro_backoff);

\end{tikzpicture}
\caption{Schematic of the collision-resolution mechanism.}
  \label{fig:auction_block_flow}
\end{figure}

\section{Regret Analysis of the Distributed Resource Allocation Mechanism}
\label{sec:regret}
This section presents the mathematical analysis of the efficiency of our resource allocation mechanism. Our main theoretical findings show that the proposed allocation mechanism guarantees a logarithmic upper bound of the network regret (see the definition in~\eqref{eq_regret}), when the network topology remains unchanged and a proper set of parameters are adopted. For ease of exposition, we first provide the main theoretical finding as follows in Theorem~\ref{Main_Theorem}:
\begin{theorem}[Main Theorem]
\label{Main_Theorem}
Assume that the numbers of time slots in the exploration, distributed auction and exploitation phases during the $j$-th epoch in Algorithm~\ref{alg:1} are $L^j_1 =L_1 = 10N^3\overline{q}^2 \Delta_{\min}^{-2}$, $L^j_2=L_2 = c N^3 \ln N$ and $L_3^j = O(2^j)$, respectively. Then, by applying Algorithm~\ref{alg:1}, the regret of the considered network after $T$ time slots is bounded by
\begin{equation}
  \label{eq_regret_bound}
  R\le O\left(N^3 \ln N \ln T \right).
\end{equation}
\end{theorem}

The proof of Theorem~\ref{Main_Theorem} can be divided into two major stages. First, the performance loss (i.e., overhead) due to the exploration and the coordination phases is expected to be logarithmic in time. Specifically, we aim to show that only a logarithmic fraction of the algorithm runtime is dedicated to exploration and coordination. Second, we need to prove that the probability of error in these two phases shrinks faster than $2^{-j}$. Since the expected regret is composed of the regrets accumulated in the exploration-auction phases and in the exploitation phase (whenever the action profile is sub-optimal), this will suffice to complete the proof. Furthermore, note that in contrast to~\cite{zafaruddin2019distributed}, our analysis inspects the expected time length used for collision resolution due to the bid quantization with finite accuracy. We also face a new framework of contention window organization due to the more complex time-frequency structure for resource division. The formal proof of Theorem~\ref{Main_Theorem} is developed in the rest of this section.

\subsection{Regret accumulated during the Exploration Phase}
\label{subsec:Pure-Exploration-Phase}
From~\eqref{eq:def:utility}, we can define the estimated utility of link $n$ under a joint action profile $\mathbf{a}$ at epoch $j$ as
\begin{equation}
  \label{eq_estimated_utility}
  \hat{u}_n^j \left(\va\right) \triangleq \hat{q}_{n,a_n}^j \mathbb{1}\left(\sum\nolimits_{m\in\cN}\mathbb{1}(a_m,a_n), 1\right),
\end{equation}
where $\hat{q}_{n,a_n}^j$ is the estimated value of the collision-free mean QoS over resource block $a_n$ by link $n$ following~\eqref{eq_update_estimate}, and $\mathbb{1}(x,y)$ is the indicator function. We define the QoS estimation errors at the end of the exploration phase of epoch $j$ as:
\begin{equation}
  \label{eq:def:QoS_est_err}
  \xi_{n,a}^j  \triangleq \left| \hat{q}_{n,a}^j -  q_{n,a} \right|,
\end{equation}
based on which, we can further define $\xi^j \triangleq \max\limits_{n\in\cN,a\in\cA} \xi_{n,a}^j$. Let $\hat{\va}^{j*}$ denote the optimal allocation derived from the estimated QoS $\hat{u}_n^j(\va)$, namely (cf.~\eqref{eq_objective}),
\begin{align}
  \label{eq_estimated_optimal_action}
    \hat{\va}^{j*}=\arg\max_{\va} \left(\hat{W}^j(\va)=\sum_{n=1}^N \hat{u}_n(\va)\right).
\end{align}
Denote as $E^j$ the event that $\hat{\va}^{j*}$ does not align with the true optimal allocation ${\va}^{*}$; i.e., $\hat{\va}^{j*}\ne {\va}^{*}$. We can obtain Lemma~\ref{lem:sufficient_utility_estimation_precision}:
\begin{lemma}
    \label{lem:sufficient_utility_estimation_precision}
    If $\xi^j \leq \xi_{\max}=\frac{3\dm}{8N}$, then, the probability of selecting a sub-optimal action profile due to using the solution in~\eqref{eq_estimated_optimal_action} is bounded by $\Pr(E^j)\leq \Pr(\xi \geq \xi_{\max})$.
\end{lemma}

\begin{IEEEproof}
It suffices to prove that $\xi \leq \xi_{\max} \Rightarrow \neg E^j$. Let $\mathbf{a}'$ be an arbitrary sub-optimal action profile. By our assumption on $q_{n,a}$, we have (see also our discussion regarding~\eqref{def:delta}):
\begin{equation}
  \label{eq:lower_boand_sub_optimality_gap_TDMA}
  W(\mathbf{a}^*) - W(\mathbf{a}') \geq \dm.
\end{equation}
Recall from Algorithm~\ref{alg:2} that the dithering value added to the local QoS estimation is sampled as $d_{n,a} \sim U \left(\left[-d_{\max}, d_{\max}\right]\right)$, where $d_{\max}=\gD_{\min}/8N$. Then, from~\eqref{eq_estimated_optimal_action} and~\eqref{eq:def:QoS_est_err}, we know that after the dithering operation in Algorithm~\ref{alg:2}, we have:
\begin{align}
  \label{eq:est_w_OFDMA_alloc_01}
  \hat{W}^j(\mathbf{a}^*) \geq {W}(\mathbf{a}^*) - \left(\xi^j + d_{\max} \right) N
\end{align}
and
\begin{align}
    \label{eq:est_w_OFDMA_alloc_02}
    \hat{W}^j \left(\mathbf{a}'\right) \leq W \left(\mathbf{a}'\right) + \left(\xi^j + d_{\max} \right) N.
\end{align}
After subtracting~\eqref{eq:est_w_OFDMA_alloc_02} from~\eqref{eq:est_w_OFDMA_alloc_01} and substituting the result into \eqref{eq:lower_boand_sub_optimality_gap_TDMA}, we obtain
\begin{equation}
    \hat{W}^j(\mathbf{a}^*) - \hat{W}^j \left(\mathbf{a}'\right) \geq \dm - 2\xi^jN - \frac{\dm}{4} \overset{\text{(a)}}{>} 0,
\end{equation}
 where (a) holds if $\xi \le \xi_{\max}$. Thus, $\xi \leq \xi_{\max} \Rightarrow \neg E^j$.
\end{IEEEproof}

To ensure that the condition $\xi \le \xi_{\max}$ in Lemma~\ref{lem:sufficient_utility_estimation_precision} holds, we need to investigate the convergence of the estimation as the number of arm samples increases for each link. For epoch $j$, let $v^j \triangleq \underset{n\in\cN,a\in\cA}{\min} v_{n,a}^{j}$ denote the number of accumulated visitations (i.e., arm-pulling) to the least-visited resource after the completion of the exploration phase. Define $v_{\min}^j \triangleq {5jN^2}/{2\Delta^2}$. Then, we can obtain the following lemma:
\begin{lemma}
\label{lem:insufficient_visitaitons}
  Let $E^j_v$ be the event for at least one link to obtain less than $v_{\min}^j\triangleq \frac{5jN^2}{2\Delta^2}$ samples of any resource block, where $\Delta$ is given by~\eqref{def:delta}. Then, $\Pr\left(E_v^j \right) = O(s^{-j})$ for some $s>2$.
\end{lemma}

\begin{IEEEproof}
  The detailed proof is presented in Appendix~\ref{app_proof_insufficient_visitaitons}.
\end{IEEEproof}

\begin{lemma}
\label{lem:insufficient_QoS_estimation_in_spite_of_sufficient_samples}
  Assume that each link collects at least  $v_{\min}^j= \frac{5jN^2}{2\Delta^2}$ samples of any resource block after epoch $j$. Then, the probability that the estimated QoS leads to a sub-optimal allocation is $\Pr\left(E^j \big| \neg E_v^j \right)= O(s^{-j})$ for some $s>2$.
\end{lemma}
\begin{IEEEproof}
  The detailed proof is presented in Appendix~\ref{app_proof_insufficient_QoS_estimation_in_spite_of_sufficient_samples}.
\end{IEEEproof}

Using Lemmas~\ref{lem:insufficient_visitaitons} and~\ref{lem:insufficient_QoS_estimation_in_spite_of_sufficient_samples}, we are able to bound the probability of the error event $E^j$ during exploration as follows:
\begin{lemma}
\label{cor:insufficient_QoS_estimation}
    After the exploration phase of epoch $j$, the optimal action profile corresponding to the estimated QoS of the channels differs from the true optimal allocation with a probability $\Pr\left( E^j \right) = O(s^{-j})$ for some constant $s>2$.
\end{lemma}

\begin{IEEEproof}
    Clearly $\Pr\left( E^j \right) \leq \Pr \left( E_v^j \right) + \Pr\left(E^j \big| \neg E_v^j \right)$. Then, from Lemmas~\ref{lem:insufficient_visitaitons} and~\ref{lem:insufficient_QoS_estimation_in_spite_of_sufficient_samples}, we have $\Pr\left( E^j \right) = O(s^{-j})$ with $s=e^{\frac{160}{81}}>2$.
\end{IEEEproof}

\subsection{Regret accumulated during the Auction Phase}
\label{subsec:Auction Phase}
We now analyze the overhead of the auction phase and its impact on the accumulated regret. In what follows, Lemmas~\ref{lem:finite_time_convergence} and~\ref{lem:blocks} analyze the convergence time of the distributed auction. Lemma~\ref{lem:optimal_convergence} provides a condition for reaching an optimal allocation upon the convergence of the auction algorithm.
\begin{lemma}
\label{lem:finite_time_convergence}
    The Distributed auction, i.e., Algorithm~\ref{alg:3} converges within at most $\frac{8N^3\overline{q}}{\dm}(1+\frac{1}{8 N})<L_2$ iterations.
\end{lemma}
\begin{IEEEproof}
    By directly applying~\cite[Lemma 3]{naparstek2013fully}, we know that link $n$ is guaranteed to be assigned a resource block within $I_n$ iterations in Algorithm~\ref{alg:3}:
    \begin{equation}
      \label{eq_number_iteration_bound}
      I_n \leq  KM + \frac{1}{\varepsilon^*}  \sum\limits_{a=1}^{N} \hat{q}_{n,a} \stackrel{(a)}{\le} 2N+\frac{N\overline{q}}{\varepsilon^*},
    \end{equation}
    where (a) follows $N=KM$ and $\hat{q}_{n,a}\leq\overline{q} + \varepsilon^*$. The simple union bound ensures that all $N$ links will be assigned a resource block within $I_{\max} \leq NI_n= N (2N+\frac{N\overline{q}}{\varepsilon^*})$ iterations. Substituting $\varepsilon^* = \frac{\dm}{8N}$ and $\frac{\overline{q}}{\gD_{\min}}\ge2$, Lemma~\ref{lem:finite_time_convergence} follows.
\end{IEEEproof}
Note that the upper bound in Lemma~\ref{lem:finite_time_convergence} is very crude. In practice, the auction algorithm typically converges much faster. Our simulations show that limiting the auction phase to be fewer than 1000 iterations for 32 links is sufficient and incurs a negligible loss in the social performance.

\begin{lemma}
\label{lem:blocks}
    The expected number of back-off blocks per auction frame (deterministic and random blocks in total) in Algorithm~\ref{alg:5} is upper bounded by $O(\ln N)$.
\end{lemma}
\begin{IEEEproof}
  The detailed proof is presented in Appendix~\ref{app_proof_blocks}.
\end{IEEEproof}

By Lemma~\ref{lem:blocks}, we know that the contention window of a single auction frame can be efficiently limited within a finite number of mini-slots. Since the CSMA mini-slot duration is typically measured in microseconds, the time length of an auction iteration (a time slot) is conveniently comparable to the length of a exploration time slot.
Now, we are ready to investigate the social optimality of the proposed auction scheme when bidding is based on the estimated QoS values.
\begin{lemma}
\label{lem:estimated_welfare_difference}
    Assume that $\mathbf{a}$ and  $\mathbf{a}'$ are two action profiles, such that for the social utilities based on the estimated QoS,
    $|\hat{W}\left(\mathbf{a}\right) - \hat{W}\left(\mathbf{a}'\right)| \neq 0$. Then with the proposed algorithm, $|\hat{W}\left(\mathbf{a}\right) - \hat{W}\left(\mathbf{a}'\right)| \geq \frac{\dm}{4}$.
\end{lemma}

\begin{IEEEproof}
    For conciseness, define $W = W\left(\mathbf{a}\right)$, $W' = W\left(\mathbf{a}'\right)$, $\hat{W} = \hat{W}\left(\mathbf{a}\right)$ and $\hat{W}' = \hat{W}\left(\mathbf{a}'\right)$.
    Then, we have
    \begin{align}
        \left| \hat{W} - \hat{W}' \right| & = \left| W - W' + \hat{W} -  W + W' - \hat{W}'  \right| \nonumber\\
        & \stackrel{(a)}{\ge}\left|W-W'\right| - \left|\hat{W}-W\right|- \left|W' - \hat{W}' \right| \nonumber\\
        & \stackrel{(b)}{\ge} \dm-\frac{3\dm}{4} = \frac{\dm}{4},
        \label{eq:estimated_welfares_difference_lower_bound}
    \end{align}
    where (a) is obtained from the reverse triangle inequality, and (b) is obtained from~\eqref{eq:lower_boand_sub_optimality_gap_TDMA} and the property of accumulated samples up to the $j$-th epoch that results in $\xi<\frac{3\dm}{8N}$; namely, following our assumption in Lemma~\ref{lem:sufficient_utility_estimation_precision}.
\end{IEEEproof}

Based on Lemma~\ref{lem:estimated_welfare_difference}, we can identify the condition for ensuring the optimal resource allocation with Algorithm~\ref{alg:3}.
\begin{lemma}
\label{lem:optimal_convergence}
    With the estimated QoS obtained from the exploration phase, the distributed auction algorithm in Algorithm~\ref{alg:3} converges to $\hat{\mathbf{a}}^*$ as long as $\varepsilon^*  \leq \dm/8N$.
\end{lemma}

\begin{IEEEproof}
This proof aims to show that a finite resolution of the contention window suffices for the distributed auction to converge. If links $n$ wins the assignment to resource $a=(k,m)$, by definition, the local $\varepsilon$-complementary slackness~\cite[Definition 1]{naparstek2013fully} guarantees (cf.~\eqref{eq_update_estimate} and~\eqref{eq:new_bid})
\begin{equation}
    \label{eq:33}
    \hat{q}_{n,a}  - b_{n,a} \geq \max_{a\in\cA} \left( \hat{q}_{n,a} - b_{n,a} \right) - \varepsilon
\end{equation}
Let the maximal bid on resource $a$ be ${b}^*_a  \triangleq \max\limits_{n\in\cN} b_{n,a}$. Since ${b}^*_a \geq b_{n,a}$, $\forall n\in\cN$, we have
\begin{equation}
    \label{eq:35}
    \max_{a\in\cA} \left( \hat{q}_{n,a} - b_{n,a} \right) \geq \max_{a\in\cA} \left( \hat{q}_{n,a} - b^*_a \right).
\end{equation}
Let $\overline{p}$ be the precision of the quantization function~\eqref{eq:new_quant} that maps the continuous bids to discrete bids. From~\eqref{eq:new_quant}, we have $\overline{p} = \overline{q} / b^*$.
When resource $a$ is assigned to link $n$, we have
\begin{equation}
    b_{n,a} > {b}^*_a  - \overline{p}.
    \label{eq:36}
\end{equation}
Substituting~\eqref{eq:36} into the left-hand side of~\eqref{eq:33} and substituting~\eqref{eq:35} into the right-hand side of~\eqref{eq:33}, we obtain
\begin{equation}
    \hat{q}_{n,a}  -  {b}^*_a \geq
    \max_{a\in\cA} \left( \hat{q}_{n,a}  - {b}^*_a \right) - (\varepsilon + \overline{p}).
    \label{eq:37}
\end{equation}

Let $\gamma \triangleq \varepsilon + \overline{p}$. \eqref{eq:37} is the global $\gamma$-complementary slackness of the original auction algorithm as defined in~\cite{bertsekas1979distributed}. With the quantization function~\eqref{eq:new_quant}, after the $\gl$ deterministic collision resolution blocks, the winner is within $\overline{p}$ from the highest bid. This is because after the $\gl$ blocks, the remaining bids all have the same quantized profit (up to $\gamma$). Together with $\varepsilon = \frac{\dm}{8N}$ we obtain $\gamma = \varepsilon+ \overline{p}= \frac{\dm}{4N}$. Let $\hat{\mathbf{a}}^*$ be the optimal joint action profile solving~\eqref{eq_estimated_optimal_action} based on the estimated QoS. Let $\tilde{\mathbf{a}}$ be the allocation obtained from Algorithm~\ref{alg:3} upon termination of the distributed auction. From~\eqref{eq:37}, we obtain
\begin{equation}
    \left|\hat{W}\left(\tilde{\mathbf{a}} \right) - \hat{W}\left(\hat{\mathbf{a}}^*\right) \right|
    \leq \gamma N = \frac{\dm}{4}.
    \label{eq:Oshri_Fully_Theorem_1}
\end{equation}
From Lemma~\ref{lem:estimated_welfare_difference} and~\eqref{eq:Oshri_Fully_Theorem_1} we have $\tilde{\mathbf{a}} = \mathbf{\hat{a}^*}$.
\end{IEEEproof}

Lemma~\ref{cor:optimal_finite_convergence} is a direct result of Lemmas~\ref{lem:finite_time_convergence} and~\ref{lem:optimal_convergence}:
\begin{lemma}
\label{cor:optimal_finite_convergence}
    If $\varepsilon^* < \frac{\dm}{8N}$, Algorithm~\ref{alg:3} for distributed auction  converges to $\hat{\mathbf{a}}^*$ within $\frac{8N^3\overline{q}}{\dm}(1+\frac{1}{8 N})$ iterations.
\end{lemma}

\subsection{Proof of the Main Theorem}
Based on the analysis of the regret accumulated from the exploration and the auction phases, we are ready to develop a formal proof of Theorem~\ref{Main_Theorem}. A duration of $T$ time slots corresponds to $J=O(log(T))$ epochs, since by the assumption $L_3^j=O(2^j)$ in Theorem~\ref{Main_Theorem},
\begin{align}
 T\geq \sum_{j=1}^{J-1} 2^{j} = (2^J-2).
\end{align}
The total regret during the exploration ($R_1$) and auction phases ($R_2$) is upper bounded by:
\begin{align}
  \label{eq_regret_exploration_auction}
  R_1 + R_2 \leq  (L_1+L_2)N \overline{q} \ln T,
\end{align}
since the regret at each time slot is upper bounded by $N\overline{q}$.

According to Lemma~\ref{cor:optimal_finite_convergence}, the exploitation phase will incur regret only if the auction phases fail to produce an allocation that is aligned with the socially optimal one based on the true QoS. By Lemma~\ref{cor:insufficient_QoS_estimation}, the probability of obtaining some sub-optimal allocation due to estimation errors in the exploration phase of epoch $j$ is upper bounded by
\begin{align}
\Pr\left(E^j\right) = O(s^{-j})
\end{align}
for some $s>2$.
Therefore, the accumulated regret from exploitation in the $J$ epochs is upper bounded by
\begin{align}
    R_3 \leq N \overline{q}  \sum_{j=1}^J (2/s)^j\leq C N \overline{q}.
\end{align}
A tighter bound using the social optimality $W^*\le N\overline{q}$ is:
\begin{align}
\label{eq:bound_exploit_regret}
    R_3 \leq W^*  \sum_{j=1}^J (2/s)^j \triangleq C W^*.
\end{align}

Summing~\eqref{eq_regret_exploration_auction} and~\eqref{eq:bound_exploit_regret}, and substituting $L_1 = 10N^3\overline{q}^2 \Delta_{\min}^{-2}$ and $L_2 = c N^3 \ln N$ therein completes the proof of Theorem~\ref{Main_Theorem}.

\section{Design toward a Practical Medium Access Control Protocol}
\label{sec:practical}
In practice, using an exponentially increasing exploitation phase is impractical. Apart from the difficulty of synchronizing links for an exponentially increasing period, this design also does not allow the links to adapt promptly to the changes in network conditions (e.g., with dynamic topology or channel states). Therefore, we suggest replacing the exponentially increasing strategy-exploitation phase with a fixed one, which contains the largest portion of each epoch.

An interesting theoretical property of the epoch structure given in Theorem~\ref{Main_Theorem} lies in the fact that the regret is mainly caused by the overhead of the exploration and auction phases, whereas the exploitation only causes an expected regret bounded by $CW^*$ following~\eqref{eq:bound_exploit_regret}. In particular, when we implement a fixed exploitation phase of $L_3$ time slots, the expected regret will be bounded by
\begin{align}
R \le \frac{L_1+L_2}{L_1+L_2+L_3} T W^*+CW^*.
\end{align}
Setting $L_3=\ga(L_1+L_2)$ results in an arbitrarily small linear regret of
\begin{align}
\label{eq:linear_regret_bound}
R \le \frac{1}{1+\ga} T W^*+CW^*,
\end{align}
for which we can choose a proper value of $\ga$ according to the channel conditions and other practical considerations.

In practice, we are interested in the efficiency of our protocol when compared to an ideal case, where the network operates under optimal allocation and is overhead-free. The efficiency can be measured by
 \begin{align}
 \mu=\frac{T W^*-\sum\limits_{t=1}^T R(t)}{T W^*}.
 \end{align}
 Simple algebra yields
$ \mu=1-\frac{\bar{R}}{W^*}$
where $\bar{R}$ is the average regret.
Using \eqref{eq:linear_regret_bound} we obtain
\begin{align}
\mu\geq  1-\frac{1}{(1+\ga)}\ -\frac{C}{T}.
\end{align}
Since $C/T$ converges to $0$, the asymptotic efficiency satisfies
 \begin{align}
    E\geq 1-\frac{1}{(1+\ga)}.
 \end{align}

Our experiments find that allocating 1-5\% of the time to channel learning and the coordination phases leads to a sufficiently accurate, fast allocation. We also find that the early stage of learning is extremely important in order to quickly adapt to network condition changes. Based on experiment results, we propose to allocate 100ms to initialize the network from scratch. This ``cold-start phase'' only consists of learning and auction phases without any exploitation. Once this initialization phase is over, the network adopts a frame structure of 5ms, which is approximately the typical coherence time of 5G channels. Within a frame, $4\gm s$ is used for channel exploration, $48 \gm s$ is used for coordination, and the rest is dedicated to exploitation. For each epoch with the deployed protocol, the links continue bidding from their bids obtained from the previous epoch.

\section{Simulation Results}
\label{sec:simulations}
This section evaluates the performance of our algorithm with numerical simulations. Throughout the simulation, the nodes of the network are placed inside a disk with a radius of 100m. External interferers are randomly placed on a ring with an inner radius of 100m and an outer radius of 200m. To create channel heterogeneity, we place a strong interferer, affecting half of the frequency bands in half of the inner disk. 20\% of the remaining resource blocks are randomly interfered by external links. We assume that the number of IoT links is known. In practice, this number can be estimated by standard techniques in a distributed manner. The main simulation parameters for the network environment is summarized in Table~\ref{tab:wireless_channel}.

\begin{figure}[!t]
  \centering
  \includegraphics[width = 0.38\textwidth]{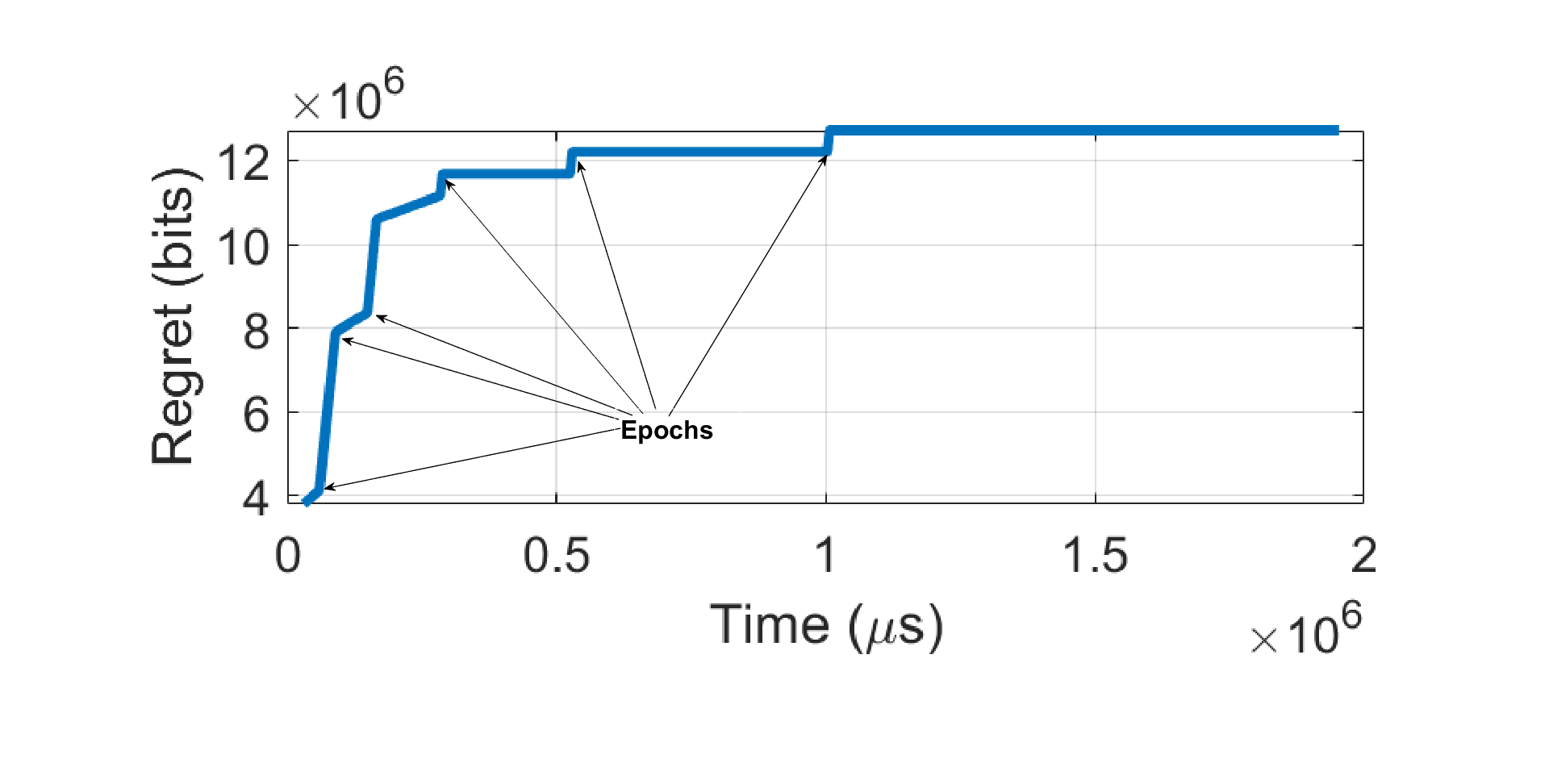}
  \caption{Regret vs. time in the considered IoT network.}
  \label{fig:Regret}
\end{figure}

\begin{figure}[!t]
  \centering
  \includegraphics[width = 0.42\textwidth]{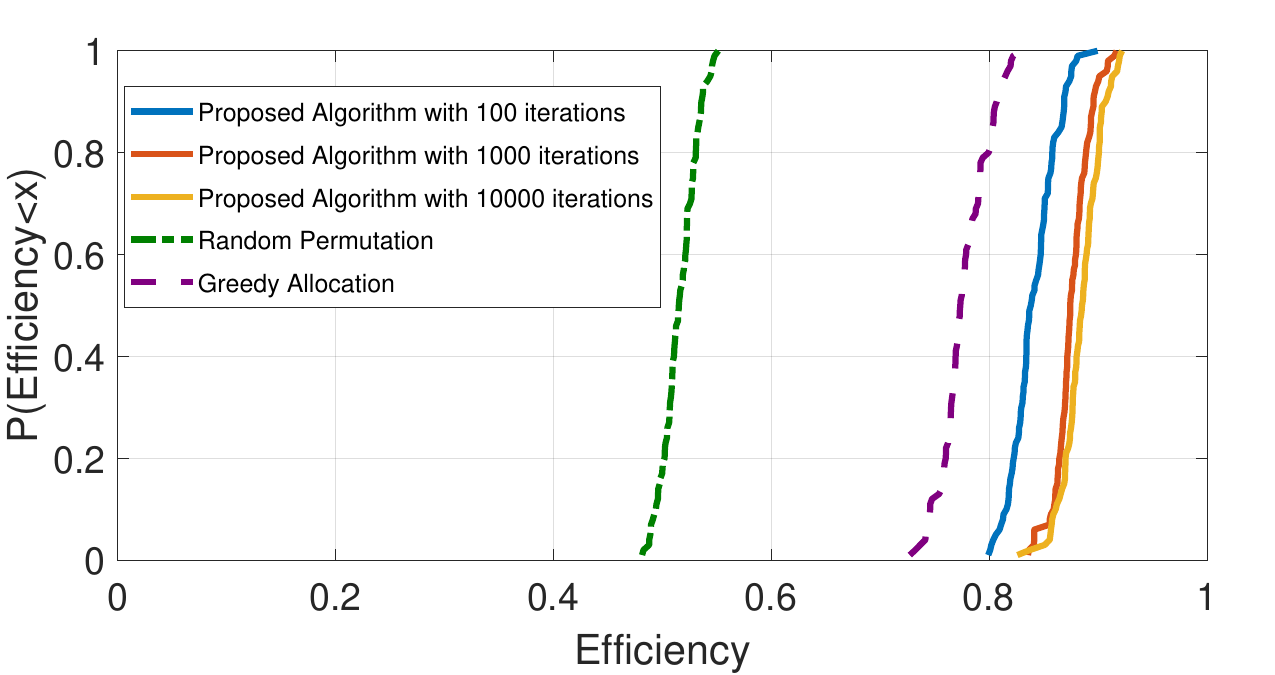}
  \caption{Monte Carlo simulation: cumulative distribution of efficiency for the proposed algorithm with different numbers of auction iterations per epoch.}
  \label{fig:eff_w_different_auction_iterations}
\end{figure}

\begin{table}[]
  \caption{Simulation parameters of the wireless environment.}
    \centering
\begin{tabular}{ | c | c | }
\hline
{\bf Network parameter} &  {\bf Value} \\
\hline
Carrier frequency & 2GHz  \\
\hline
Total bandwidth & 40MHz  \\
\hline
Sub-channel bandwidth & 5MHz  \\
\hline
Path loss exponent $\alpha$ & 4 \\
\hline
No. of uniformly distributed tap delays & 7 \\
\hline
Rayleigh Variance & $10^{-2}$ \\
\hline
Log-mean of shadowing & 0 \\
\hline
Log-variance of shadowing & 0.01 \\
\hline
Transmit power & 1mW \\
\hline
Noise power spectrum density & $-174$dBm/Hz \\
\hline
Interference power spectrum density & $-57$dBm/Hz \\
\hline

\end{tabular}
\label{tab:wireless_channel}
\end{table}
{
The standard multipath propagation model is given by the well-known Saleh-Valenzuela model \cite{saleh1987statistical,meijerink2014physical}: 
\begin{align}
    h_{ij}(t)=C d_{i,j}^{-\ga/2} \sum_{\ell=1}^L h_{\ell}\gd(t-\gt_\ell) 
\end{align}
where the variance of the multipath component $h_{\ell}$ decays with the delay, i.e., it is a Rayleigh fading with variance decreasing with $\gt_{\ell}$. Typically, in standards (for point-to-point models) these are tabulated. However, in the context of a network, these should be randomized. To that end, we defined $L$ as in Table I, where the number of random paths is $7$ to fit the strongest multipath models in standards. for each link, $i,j$ we have $d_{i,j}$ defined by the network topology as the distance between the two stations $i,j$. $\gt_{\ell}$ is the additional delay of each path compared to the LOS. We picked $0\le \gt_{\ell} \le \gt_{\max}$, where $\gt_{\max}$ satisfies 
\begin{align}
   \left(1+\frac{c\gt_{\max}}{d_{i,j}}\right)^{-\ga/2} = 0.1.  
\end{align}

The  tap $h_\ell$ was constructed by picking a unit complex-normal fading coefficient and multiplying it by $(1+\frac{c\gt_{\ell}}{d_{i,j}})^{-\ga/2}$.  This makes the decay of each path consistent with the path loss model and the selection of $\gt_{\ell}$, since, the total distance of the $\ell$'th path is $d_{i,j}+c\gt_{\ell}$. Therefore, the energy loss along this path is expected to be  $(d_{i,j}+c\gt_{\ell})^{-\ga}=d_{i,j}^{-\ga}(1+\frac{c\gt_{\ell}}{d_{i,j}})^{-\ga}$. The taps are selected independently. This model is then translated into the frequency domain to obtain the received power. 
On top of this model, The channel impulse response was multiplied by a standard log-normal shadowing coefficient with the parameters given in Table I.  
The same model was used to calculate the effect of external interference on each station. }
Our first experiment depicts the regret evolution with an exponentially increasing number of time slots during the exploitation phase as shown in Figure~\ref{fig:Regret}, where a fully-stationary network with fixed topology is simulated. The allocation is learned in 6 epochs with 1000 estimation time slots and 400 auction iterations in each epoch. Note that although the theoretical convergence bound for the auction iteration is large, we observed that terminating the auction after 1000 iterations suffices to achieve an efficient allocation (see Figure~\ref{fig:eff_w_different_auction_iterations}, where an efficiency of 85\% can be reached with a probability of 0.8 with truncated iterations). The details of the parameter setting for the simulation are given in Table~\ref{tab:regret_para}\footnote{Note that in this experiment we use an exponentially growing exploitation to demonstrate the theoretical result.}. Clearly, after the third epoch, the regret becomes 0 during the exploitation phase and the accumulated regret is due solely to exploration and auction.

\begin{figure}[!t]
  \centering
  \includegraphics[width = 0.38\textwidth]{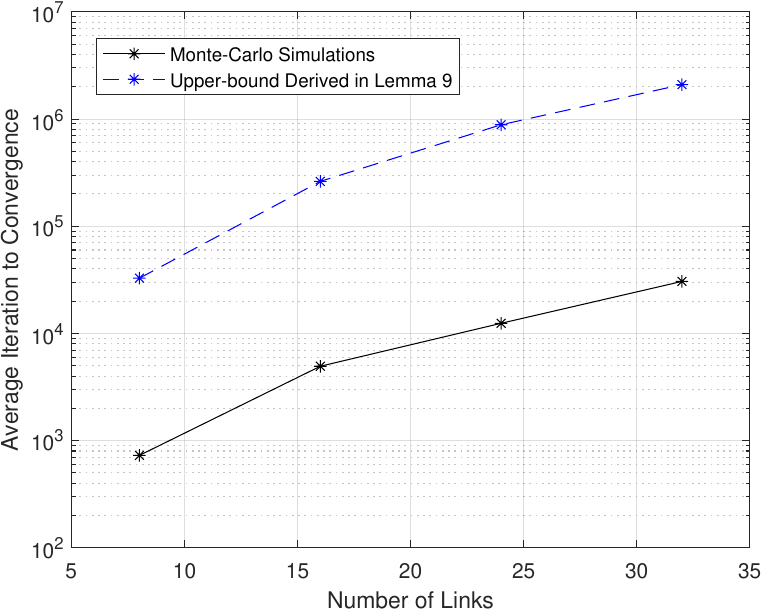}
  \caption{Average iterations to convergence vs.network size $N$ ($K=8$).}
  \label{fig:iter_vs_regret_varying_UE_num}
\end{figure}

{  Our second experiment investigates the impact of the network size and the number of links on the convergence speed and efficiency of the proposed algorithm. It is worth noting that although Lemma~\ref{cor:optimal_finite_convergence} provides a loose bound of complexity for the auction algorithm in terms of link number $N$, i.e., $O(N^3)$, in practice, the number of total auction iterations needed for convergence is much smaller than that. This is supported by our simulation results in Figure~\ref{fig:iter_vs_regret_varying_UE_num}, where the average number of iterations required for convergence is significantly smaller than the value estimated using the bound.}

\begin{table}[!t]
  \caption{Parameters of the MAC protocol for the first experiment.}
\begin{center}
\begin{tabular}{ | l | c |}
\hline
{\bf MAC super-frame structure parameter} &  {\bf Value} \\
\hline
Number of epochs & 6 \\
\hline
Exploration time per epoch &  $4 ms$ \\
\hline
Expected auction time per epoch & $12ms$ \\
\hline
Number of possible discrete bids & 256 \\
\hline
\end{tabular}
\end{center}
\label{tab:regret_para}
\end{table}

\begin{figure}[!t]
  \centering
  \includegraphics[width = 0.45\textwidth]{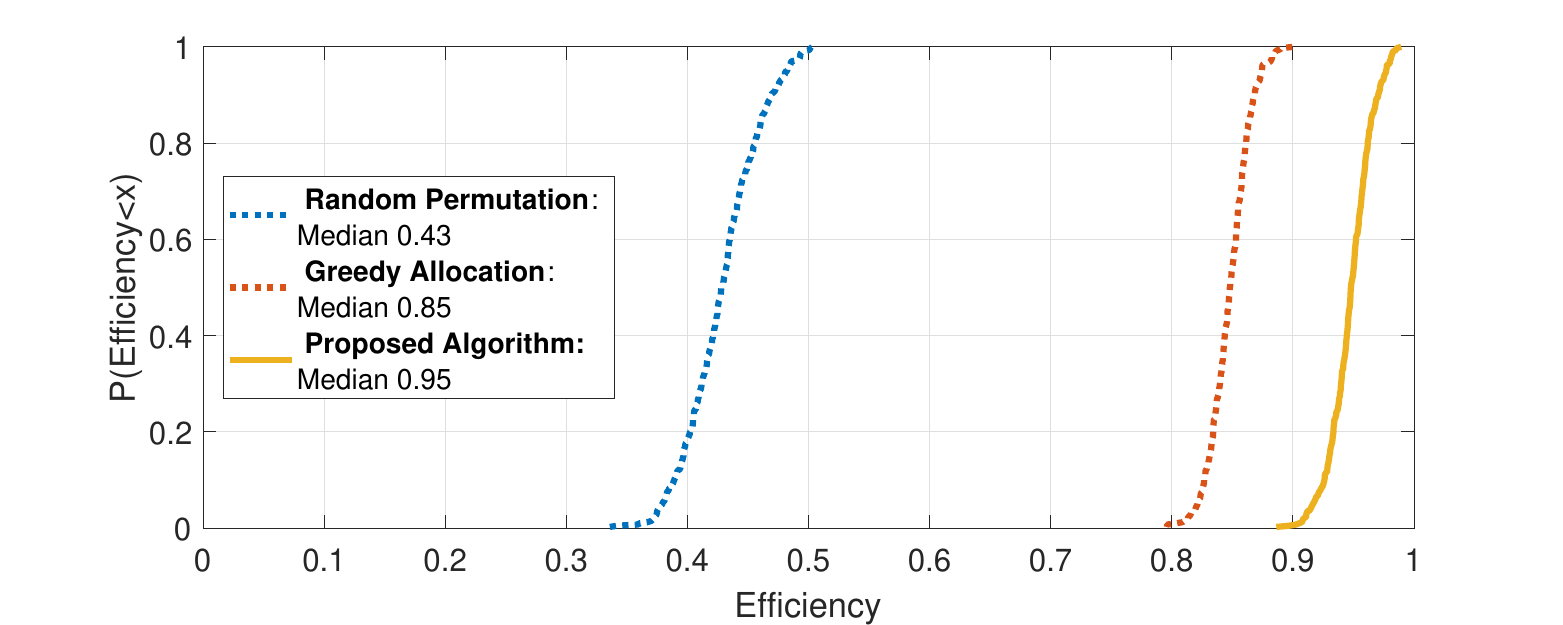}\vspace{-1mm}
  \caption{Cumulative distribution of efficiency for different algorithms after 100 epochs of fixed-size in a static network of $N=32$ and $K=8$.}
  \label{fig:eff_infinite}
\end{figure}

\begin{figure}[!t]
  \centering
  \includegraphics[width = 0.45\textwidth]{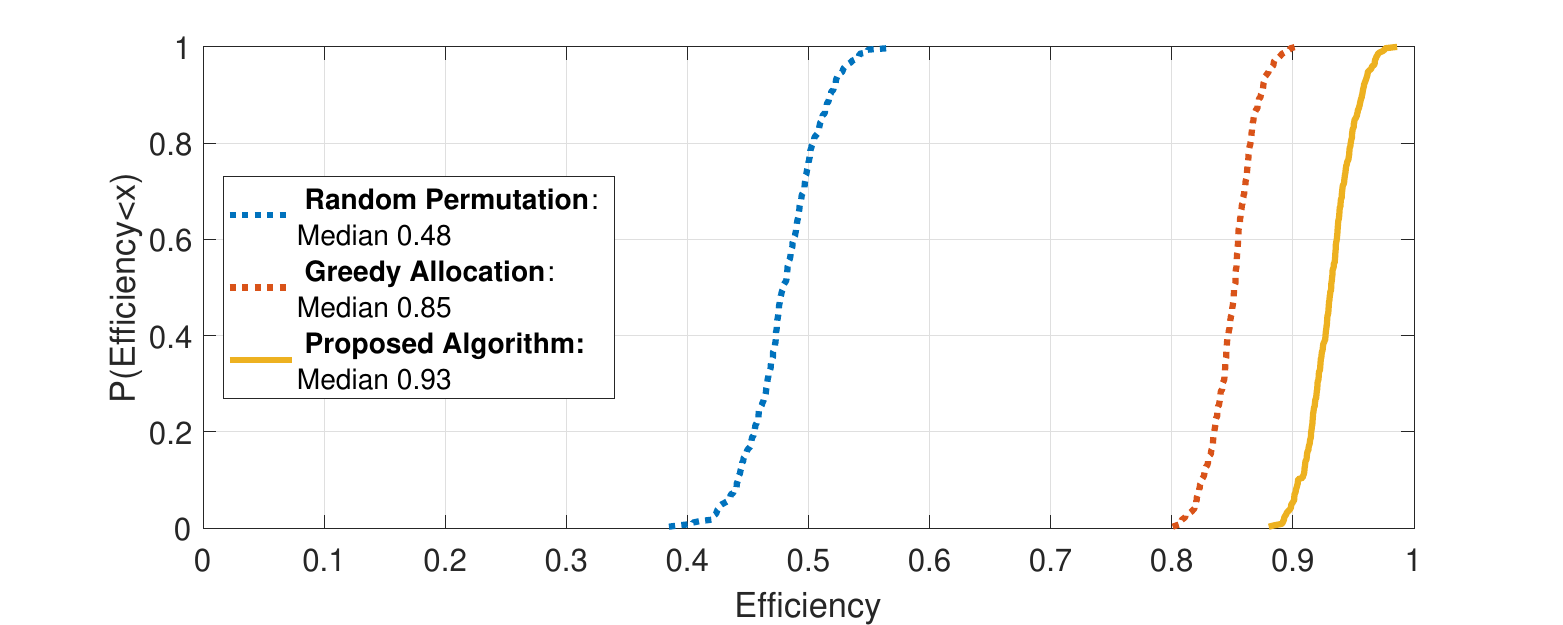}\vspace{-1mm}
  \caption{Cumulative distribution of efficiency for different algorithms after 100 epochs of fixed-size in a dynamic networks of $N=32$ and $K=8$.}
  \label{fig:eff_finite}
\end{figure}

The third experiment is set up in a dynamic environment with varying channel conditions. In this case, a cold-start stage of learning and auction is adopted, followed by repeated epochs of three fixed-size phases. Tables~\ref{tab:setup_mac} and~\ref{tab:steady_state_parameters} present the related parameters. Figures~\ref{fig:eff_infinite} and~\ref{fig:eff_finite} demonstrate the performance of this protocol with a network cold-start stage of $100$ms followed by 100 epochs of a fixed time length ($5ms$). In Figures~\ref{fig:eff_infinite}, and Figure~\ref{fig:eff_finite} the cumulative distribution for the allocation efficiency is presented for a static radio environment and a dynamic environment (with a channel coherence time of $5ms$), respectively. The performance in terms of efficiency is measured with respect to the {\bf optimal centralized allocation derived using the Hungarian algorithm based on the true channel state information}. The performance of greedy allocation (i.e., using stable matching) and random allocation/permutation are presented for comparison, both using the same collision avoidance technique to derive their allocation (see also~\cite{leshem2011multichannel}), 
{
Random allocations are used when channel state information is not available at all stations and no signaling protocol as proposed here is used. 
The greedy algorithm (which is equivalent to a stable matching) in this context has also been proposed \cite{leshem2011multichannel}, where it was shown to have excellent performance over i.i.d Rayleigh fading channels (but also there, the learning aspects were not discussed). It is simulated in the current context of external interference. We believe these two techniques are the baseline for comparison for any other allocation. 
}

\begin{table}[!t]
  \caption{Parameters in the ``cold-start'' phase.}
\centering
\begin{tabular}{ | l | c | l | c| }
\hline
{\bf Parameters} &  {\bf Value} & {\bf Parameters} &  {\bf Value} \\
\hline
Exploration duration & $85ms$ & Initial $\varepsilon$ & 1 \\
\hline
Auction duration & $15ms$ & Final $\varepsilon$ & $1/32$ \\
\hline
Maximum auction iterations & 500 & $\beta$ & 4\\
\hline
$\varepsilon$-scaling parameter  $\zeta$ & 0.9808  & & \\
\hline
\end{tabular}
\label{tab:setup_mac}
\end{table}

\begin{table}[!t]
  \caption{Parameters for learning and exploitation in one epoch.}
\begin{center}
\begin{tabular}{ | l | c | l | c |}
\hline
{\bf Parameters} &  {\bf Value} & {\bf Parameters} &  {\bf Value} \\
\hline
Epoch duration  & 5ms & Initial $\varepsilon$  & 1/32 \\
\hline
Exploration time per epoch &  $50 \mu s$ & Final $\varepsilon$ & 1/32 \\
\hline
Expected auction phase duration  & $ 200 \mu s$ & $\zeta$ & 1\\
\hline
Exploitation phase  & $4750 \gm s$ & & \\
\hline
\end{tabular}
\end{center}
\label{tab:steady_state_parameters}
\end{table}

Figures~\ref{fig:eff_infinite} and~\ref{fig:eff_finite} show that the proposed protocol achieves more than $90\%$ efficiency in both the static and dynamic network environments. Our algorithm clearly outperformed the random allocation, which achieved less than 50\% efficiency. While the greedy allocation achieves in expectation 85\% efficiency in both the static and dynamic network environment setting, our algorithm achieved 93\% efficiency in the dynamic environment and 95\% in the static environment. The proposed algorithm also maintains a $5\%$ outage rate at around 90\% efficiency for both the static and dynamic settings, while the greedy algorithm achieves at best $80\%$ efficiency relative to the optimal centralized solution.

\begin{figure}[!t]
    \centering
\includegraphics[width = 0.35\textwidth]{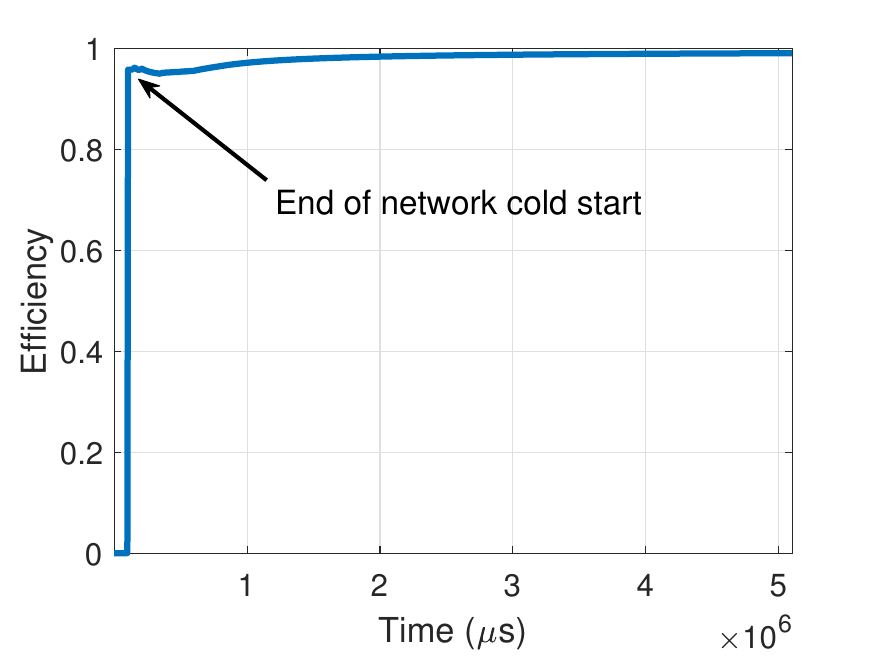}\vspace{-1mm}
\caption{Efficiency evolution of the proposed protocol with a cold-start stage and 100 epochs.}
\label{fig:Efficiency_Semilogx_Full_Length}
\end{figure}

\begin{figure}[!t]
    \centering
\includegraphics[width = 0.35\textwidth]{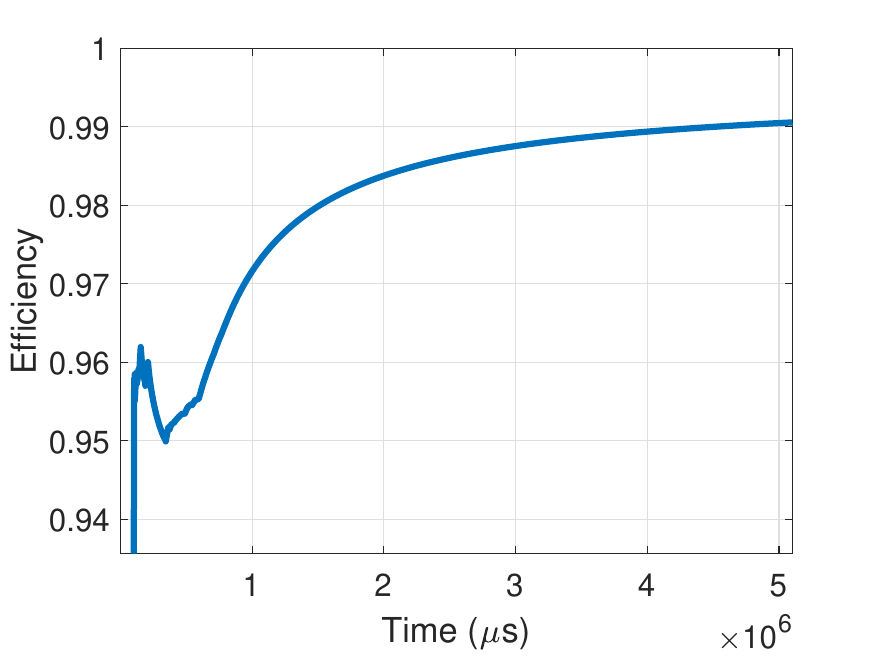}\vspace{-1mm}
\caption{Efficiency evolution in 100 epochs of steady state.}
\label{fig:Efficiency_Semilogx_Zoom_In}
\end{figure}

Figures~\ref{fig:Efficiency_Semilogx_Full_Length} and \ref{fig:Efficiency_Semilogx_Zoom_In} show the evolution path of the efficiency of the proposed algorithm for a single dynamic network with the same network settings. At most $1\%$ of the total algorithm runtime is dedicated to learning, so that the maximal efficiency\footnote{If the auction converges early, the learning process will take less time and the efficiency will be even higher.} is around $99\%$. The figures show that the efficiency converges to this value very quickly.
Figure~\ref{fig:Efficiency_Semilogx_Full_Length} depicts the complete evolution including the network cold-start stage and 100 fixed-length epochs. It illustrates the rapid learning trend during the network cold-start stage, shortly after which, the algorithm converges to the maximum possible efficiency of $99\%$. Figure~\ref{fig:Efficiency_Semilogx_Zoom_In} zooms in on the steady-state part of Figure~\ref{fig:Efficiency_Semilogx_Full_Length}. Note that because of the dynamics of the channels, there are situations where efficiency is reduced until the protocol re-converges to the optimal allocation.

\section{Conclusion}
\label{sec:conclude}
We presented an algorithm for distributed spectrum allocation in dense D2D networks, that requires no direct communication between the ad-hoc links. The algorithm relies solely on the local CSI estimation and employs a distributed auction mechanism that is enabled by single-channel opportunistic carrier sensing. Time-sharing is introduced to relieve network congestion. The proposed algorithm achieves implicit collaboration for access scheduling through epoch-based learning for the optimal time-frequency configuration with a small overhead. With an exponentially growing window of strategy exploitation, the proposed algorithm achieves a sublinear regret bound of $O(\ln T)$. Comparatively, with a fixed window of exploitation, our proposed protocol achieves an exponentially decaying probability of obtaining a sub-optimal allocation profile with an arbitrarily low overhead. The numerical simulations demonstrate that our protocol significantly outperforms the greedy allocation using stable matching and the random allocation scheme when applied to 5G channels.
\label{sec:conclusions}

\appendices
\section{Proof of Lemma~\ref{lem:insufficient_visitaitons}}
\label{app_proof_insufficient_visitaitons}
From Algorithm~\ref{alg:2}, a link independently chooses a resource block to send its pilots. Let $E_{n,a}$ denote the event that link $n\in\cN$ sees no collision upon its selected action $a=(k,m)$ at a given time slot. Recall that $KM=N$, then,
\begin{equation}
  \label{eq_no_collision_probability}
  \Pr(E_{n,a}) = \frac{1}{N}\left(1-\frac{1}{N}\right)^{N-1}.
\end{equation}

For link $n\in\cN$, let $H_{n,a}$ be the number of event $E_{n,a}$ occurring in  $\ell$ independent trials. $\ell$ corresponds to the number of exploration slots up to the $j$-th epoch. The standard form of Hoeffding's inequality for Bernoulli random variables given the success probability $\Pr(E_{n,a})$ and the number of trials $\ell$ is:
\begin{equation}
\label{eq:hoeffding1}
  \Pr\left(H_{n,a}\leq (\Pr(E_{n,a})-\eta)\ell\right) \leq \exp(-2\ell\eta^2 ).
\end{equation}
Based on our assumption of $L_1 = 10N^3 \overline{q}^2\Delta_{\min}^{-2}$ in Theorem~\ref{Main_Theorem}, we have $\ell=10j N^3\Delta^{-2}$ from (\ref{def:delta}). Then,~\eqref{eq:hoeffding1} becomes
\begin{equation}
\label{eq:hoeffding2}
  \Pr\left(H_{n,a}\leq (\Pr(E_{n,a})-\eta)10 \frac{j N^3}{\Delta^2}\right) \leq e^{-20\frac{j N^3}{\Delta^2}\eta^2}.
\end{equation}
Let $v^j_{\min}=(\Pr(E_{n,a})-\eta)10j N^3\Delta^{-2}$. Then, we have
\begin{equation}
\label{eq_epsilon_value}
  \eta = \Pr(E_{n,a}) - \frac{v^j_{\min}}{10jN^3 \Delta^{-2}}.
\end{equation}
Substituting~\eqref{eq_epsilon_value} into~\eqref{eq:hoeffding2}, we have
\begin{align}
\label{eq:hoeffding3}
  & \Pr(H_{n,a}\leq v^j_{\min}) \nonumber \\
  & \leq \exp\left(-20 j N^3\Delta^{-2} \left(\Pr(E_{n,a}) - \frac{v^j_{\min}}{10jN^3 \Delta^{-2}}\right)^2\right).
\end{align}
Substituting~\eqref{eq_no_collision_probability} and $v^j_{\min} =\frac{5jN^2}{2\Delta^2}$ into~\eqref{eq:hoeffding3} yields
\begin{align}
\label{eq:hoeffding4}
  & \Pr(H_{n,a}\leq v_{\min})  \nonumber \\
  & \leq\exp \left(-20jN\Delta^{-2} \left(\left(1 - \frac{1}{N} \right)^{N-1} - \frac{1}{4}\right) ^2 \right).
\end{align}
Since $\lim\limits_{N\rightarrow\infty}(1-\frac{1}{N})^{N}=e^{-1}$, we have $\left(1-\frac{1}{N}\right)^{N-1} - \frac{1}{4} \geq e^{-1}-\frac{1}{4} > \frac{1}{9}$. Then, from~\eqref{eq:hoeffding4} we obtain
\begin{align}
\label{eq_relaxed_hoeffding}
  \Pr(H_{n,a}\leq v_{\min}) < \exp\left(-20 j N\Delta^{-2} /81\right).
\end{align}

Let $E_v^j$ denote the event that at least one link accumulates fewer than $v^j_{\min}$ samples over some of the $MK$ resource blocks by the end of epoch $j$. Then, we have
\begin{eqnarray}
\label{eq_union_bound}
  \Pr(E_v^j) = \bigcup\limits_{n\in\cN}\bigcup_{a\in\cA} \Pr(H_{n,a}\leq v^j_{\min})
  \stackrel{(a)}{<}  N^2 e^{-\frac{20}{81}jN\Delta^{-2}},
\end{eqnarray}
where (a) is obtained using the union bound and the condition $MK=N$. Then, it is sufficient to have
\begin{equation}
\label{eq:exp_const_big}
  \frac{20}{81} N\Delta^{-2} > \ln(2),
\end{equation}
to complete the proof. Since by our assumption $N\ge2$ and $\gD^{-1}={\overline{q}}/{\gD_{\min}}\geq 2$, the error probability satisfies
\begin{equation}
\Pr(E_v^j) = O(s^{-j}),
\end{equation}
where $s=e^{\frac{160}{81}}>2$.

\section{Proof of Lemma~\ref{lem:insufficient_QoS_estimation_in_spite_of_sufficient_samples}}
\label{app_proof_insufficient_QoS_estimation_in_spite_of_sufficient_samples}
By our assumption in Section~\ref{sec:system}, the samples successfully collected for a resource block during an epoch are i.i.d. and bounded by $[0, \overline{q}]$. For a total number of $v_{n,a}$ samples collected by link $n$ on action $a$, we have
\begin{equation}
  \label{eq_estimated_qos}
  \hat{q}^j_{n,a}=\frac{1}{v_{n,a}}\sum_{t=1}^{v_{n,a}}q^t_n(a).
\end{equation}
From the proof of Lemma~\ref{lem:sufficient_utility_estimation_precision}, we have
\begin{align}
  \label{eq:13}
  \Pr\left( E^j \big| \neg E_v^j \right) & = \bigcup\limits_{n\in\cN}\bigcup_{a\in\cA}\Pr \left( \vert \hat{q}^j_{n,a} -q_{n,a} \vert > \xi_{\max} | \neg E_v^j\right) \nonumber \\
  & \stackrel{(a)}{\le} \bigcup\limits_{n\in\cN}\bigcup_{a\in\cA} 2\exp\left(-\frac{2v^2_{n,a}\xi^2_{\max}}{v_{n,a}\overline{q}^2} \Bigg| \neg E_v^j\right)  \nonumber \\
  & \stackrel{(b)}{\le} 2N^2\exp\left(-2\xi_{\max}^2 v_{\min}^j /\overline{q}^2 \right),
\end{align}
where (a) is obtained following Hoeffding's inequality, and (b) is obtained based on the union bound and the condition $\forall n, a: v_{n,a}\ge v^j_{\min}$ for $\neg E_v^j$. Substituting $v_{\min}^j$, $\xi_{\max}$ and $\Delta$ (see also~\eqref{eq:def:QoS_est_err} and~\eqref{def:delta}, respectively) into \eqref{eq:13} we obtain
\begin{align}
  \label{eq:proof_bound_of_error_exploration}
    \Pr\left( E^j \big| \neg E_v^j \right)
    = 2 N^2 \exp\left( -\frac{45 j}{64}\right).
\end{align}
Then, with $s=e^{\frac{45}{64}}>2$, $\Pr(E^j \big| \neg E_V^j) = O(s^{-j})$.

\section{Proof of Lemma~\ref{lem:blocks}}
\label{app_proof_blocks}
By~\eqref{eq_number_of_digits}, the number of deterministic back-off blocks is bounded by $\gl=O(\ln N)$. Hence, it suffices to provide the bound on the number of the random back-off blocks. We consider a general case where $\ell_a\ge 2$ links offer the same highest quantized bid for resource block $a$ in one iteration. Thus, they enter the random back-off period with repeated binomial trials (see also Figure~\ref{fig:Auction_Phase}). Assume that $\ell_{a,1}$ links out of the total number of $\ell_a$ links choose to transmit in the first block. If $\ell_{a,1}\ge2$, they will continue to randomly back-off\footnote{If $\ell_{a,1}=0$, all the links also continue to back-off in the next mini-slot.} in the next block. Therefore, the expected number of links competing in the next block is:
\begin{align}
\label{eq:17}
  \Ep\left\{\ell_{a,2}\right\}  & = \sum_{\ell_{a,1}= 2}^{\ell_a} \binom{\ell_a}{\ell_{a,1}} \frac{\ell_{a,1}}{2^{\ell_a}} + \frac{\ell_a}{2^{\ell_a}}  \nonumber \\
  & \leq \ell_a \left( \frac{1}{2}+2^{-\ell_a} \right) \stackrel{\ell_a\ge2}{\leq} \frac{3}{4} \ell_a.
\end{align}
Let $C_1= \ln(4/3)$, and $\ell_{a,B}$ be the number of random back-off blocks (each containing one mini-slot for back-off and one mini-slot for collision notification) needed to determine a winner in the contention. From \eqref{eq:17}, we need $(\frac{3}{4})^{\ell_{a,B}}\ell_a\le 1$. Then, if $\ell_a\ge2$, we need
\begin{equation}
\label{eq:18}
  \Ep \left\{ \ell_{a,B} \right\} \ge  C_1\ln\ell_a.
\end{equation}
to terminate the random backoff. Let $t$ be the number of blocks needed after $C_1\ln\ell_a$ blocks for winner determination. With the same technique of deriving \eqref{eq:17}, we have
\begin{equation}
\label{eq:18_plus}
  \Pr \left( \ell_{a,C_1\ln\ell_a+t} > 1\right) \le  \left(\frac{3}{4}\right)^t,
\end{equation}
which shrinks exponentially. Thus, the expected number of back-off blocks over resource block $a$ is upper bounded by
\begin{equation}
\label{eq:18_converge}
  \Ep \left\{ \ell_{a,B} \right\} \le C_1\ln\ell_a + \sum_{t=1}^{\infty} (C_1\ln\ell_a+t)  \left(\frac{3}{4}\right)^t \le C_1\ln\ell_a + C_2,
\end{equation}
where $C_2$ is a constant.

The number of links choosing resource block $a$ in a given iteration is a binomial random variable $Z_a \sim B(N, \frac{1}{N})$. Naturally, $\ell_a \leq Z_a$. According to \cite[Corollary A.1.10]{alon2004probabilistic}, we have the following inequality:
\begin{equation}
\label{eq:Noga_Alon}
  \Pr\left(\ell_a \geq x \right) < \exp \left( x - \left( x+1 \right) \ln \left( 1+x \right) \right) < \left( \frac{e}{x}\right)^x.
\end{equation}
Let $x = \ln^2 N$. Then, for $N \geq 12$, $\left( \frac{e}{x}\right)^x < N^{-2}$. From~\eqref{eq:Noga_Alon},
\begin{equation}
  \label{eq:probability_of_hight_load_bound}
  \Pr(\ell_a \geq \ln^2 N) < N^{-2}.
\end{equation}
From~\eqref{eq:18_converge} and~\eqref{eq:probability_of_hight_load_bound}, the expected number of random back-off blocks over a resource block $a=(k,m)$ is upper bounded by
\begin{align}
  \label{eq:20}
  \Ep \left\{ \ell_{a,B} \right\} & = \Ep \left\{  \ell_{a,B}|\ell_a \geq \ln^2 N  \right\} + \Ep \left\{ \ell_{a,B}|\ell_a < \ln^2 N \right\} \nonumber \\
  & \leq N^{-2} \left(C_1\ln N + C_2\right) + C_1\ln \left( \ln^2 N \right) + C_2,
\end{align}
which completes the proof.

\end{document}